\documentclass[%
 reprint,
superscriptaddress,
 amsmath,amssymb,
 aps,
]{revtex4-2}
\usepackage [latin1]{inputenc}
\usepackage{graphicx}
\usepackage{dcolumn}
\usepackage{bm}
\usepackage{subcaption}
\usepackage{float}
\usepackage{xcolor}
\usepackage[hidelinks]{hyperref}

\begin{document}

\preprint{APS/123-QED}

\title{Wake Interference Effects on Flapping Dynamics of Elastic Inverted Foil}

\author{Aarshana Parekhi}
\email[Corresponding author. Email address: ]{arparekh@student.ubc.ca}
\affiliation{Department of Mechanical Engineering, The University of British Columbia, Vancouver, BC Canada V6T 1Z4}

\author{Rajeev K. Jaiman}
\email[Email address: ]{rjaiman@mech.ubc.ca}
\affiliation{Department of Mechanical Engineering, The University of British Columbia, Vancouver, BC Canada V6T 1Z4}

\date{\today}

\begin{abstract}
Using high-fidelity simulations, we study the self-induced flapping dynamics of an inverted elastic foil when it is placed in tandem with a stationary circular cylinder. The effect of wake interference on the inverted foil's coupled dynamics is examined at a fixed Reynolds number ($Re$) as a function of non-dimensional bending rigidity ($K_{B}$) and the structure to fluid mass ratio ($m^{*}$). Our results show that there exists a critical $K_{B, Cr} = 0.25$, above which the downstream foil is synchronized with the unsteady wake, and the cylinder controls the flapping response and the wake vortex dynamics. During synchronization, two additional flapping modes namely, small and moderate amplitude flapping mode are observed as a function of decreasing $K_{B}$. Below $K_{B, Cr}$, the downstream foil undergoes self-induced large-amplitude flapping (LAF) similar to an isolated foil counterpart. When the dynamics of the downstream foil are analyzed for a range of $m^{*}$, we can characterize the response dynamics into two regions, namely low and high sensitivity. The high sensitivity region is observed when the dynamics are controlled by the cylinder oscillations, i.e., for foils with high stiffness. In this regime, the foil dynamics negatively correlate to $K_{B}$ and $m^{*}$. The low sensitivity region is observed when the downstream foil is no longer synchronized with the wake and undergoes an LAF response, with dynamics that are weakly correlated to $K_{B}$. A new non-dimensional parameter is proposed that combines the effect of the foil's inertia and elastic forces and can capture the foil's response when it is subjected to wake interference effects. The findings from this study aim to generalize our understanding of the self-induced flapping dynamics of inverted foils in an array configuration and have relevance to the development of inverted foil-based renewable energy harvesters.
\end{abstract}

\maketitle
\section{\label{sec:introduction}Introduction\protect}
A cantilevered flexible foil or panel exhibits self-induced flapping oscillations when immersed in a flowing fluid. Particularly, flexible foils anchored at the trailing edge (i.e., inverted foils) demonstrate large amplitude oscillations across a spectrum of system parameters. This complex fluid-structure phenomenon is not only noteworthy for its prevalence in nature but also for its wide-ranging engineering applications, including but not limited to paper printing  \cite{Watanabe2002AFlutter} and nuclear plate assemblies \cite{Guo2000StabilityFlow}. Of specific interest is its application in small-scale renewable energy harvesting. Through flapping motion, these thin cantilevered flexible foils can effectively capture wind/marine hydro-kinetic energy, which can be subsequently converted into electrical energy via coupling with piezoelectric materials \cite{ Allen2001EnergyEel,kim2013flapping,Orrego2017HarvestingFlag,ribeiro2021wake}. The flexible-foil based energy harvesters generate power in micro-Watt to milli-Watt scale \cite{Orrego2017HarvestingFlag} suitable powering devices such as remote sensors, data transmitters, and small-scale portable electronics \cite{Mathuna2008EnergyNetworks,erturk2010energy,Erturk2011PiezoelectricHarvesting,ribeiro2021wake}.  Due to the self-sustaining nature of oscillations, with meticulous calibration and active control, these flexible foils could continuously generate power through synchronized fluid-structure coupling.

In a uniform flow stream, a flexible foil is classified based on how the foil is cantilevered.  The first is the conventional configuration wherein the foil is cantilevered at the leading edge, with its trailing edge left free to oscillate. When subject to fluid flow, the conventional foil performs flapping motion for flow velocities greater than a critical velocity \cite{taneda1968waving}. Instability is induced when the structure cannot dissipate the kinetic energy of the fluid. Flapping motion is observed due to the structure's inherent fluid-elastic instability resulting from positive feedback between the foil's inertia force and its elastic force and fluid force \cite{Argentina2005Fluid-flow-inducedFlag}.  A large number of energy harvesting models have been proposed based on the idea of utilizing conventional foils to harness fluid kinetic energy \cite{Allen2001EnergyEel,Akcabay2012HydroelasticFlow,Michelin2013EnergyFlows}. However, these studies showed that conventional foils lose their stability at very low foil flexibility values, which correspond to very high flow velocities that are not favorable for energy harvesting. Moreover, the conventional foils vibrate with small amplitudes, and their instability dynamics are sensitive to design conditions; structure's mass \cite{Zhang2000FlexibleWind,Shelley2005HeavyWater, Shelley2011FlappingFlows,Connell2007FlappingStream}, flow viscosity \cite{Jaiman2014AddedChannel,Liu2014AEffects}, and foil aspect ratio \cite{Eloy2007FlutterPlate,Eloy2008AeroelasticFlow} to name a few. In terms of energy harvesting efficiency, for high structure-to-fluid mass ratios($m^*$) of $O(10)$ conventional foils can have efficiency up to $10 - 12 \%$, but for low structure-to-fluid mass ratios of $O(10^{-3})$, the efficiency is very low \cite{Akcabay2012HydroelasticFlow} and \cite{Michelin2013EnergyFlows}.

The second configuration is of an inverted foil, in which the foil is cantilevered at the trailing edge, and the leading edge is left free to oscillate. In this configuration, the incoming fluid impinges the foil at the free end. Consequently, the inverted foil is susceptible to instability at much lower flow velocities \cite{Zhang2000FlexibleWind}. The instability dynamics and mechanisms in this configuration are also very different compared to the conventional foil. If we consider an inverted foil of a given mass and stiffness under steady uniform flow. This foil experiences fluid loading as the flow moves around the surface, and in an equilibrium state, these forces are balanced by the structural elastic forces. As flow velocity is increased, it first undergoes static divergence \cite{kim2013flapping,Gurugubelli2015Self-inducedFlow}, and eventually, flapping instability is induced when foil can no longer dissipate the fluid kinetic energy.  Three distinct flapping regimes are observed for inverted foils: (i) large-amplitude flapping (LAF) regime, (ii) deflected flapping, and (iii) flipped flapping regime.

In the LAF regime,  the foil undergoes limit cycle oscillations for a range of velocities with amplitudes almost five times larger than conventional foils \cite{kim2013flapping,Gurugubelli2015Self-inducedFlow}. These large amplitude oscillations are sustained by the combined effect of periodic vortex shedding at the leading edge and the foils inherent fluid elastic-instability \cite{Gurugubelli2019LargePeriodicity,Goza2018GlobalFlapping,Tavallaeinejad2018NonlinearFlow,Tavallaeinejad2020InstabilityFlow}. Unlike their traditional counterparts, large changes in a mass ratio only influence the flapping frequency, vortex shedding mode, and transition from LAF mode to deflected flapping mode. While the stability boundary of transverse vibration amplitudes is only weakly influenced by the change in $m^{*}$ \cite{Gurugubelli2015Self-inducedFlow, Shoele2016EnergyFlag,Tang2015DynamicsFlow}. Experimental and numerical studies conducted at high Reynolds number $Re \sim O(10^4)$  \cite{kim2013flapping, Sader2016Large-amplitudeVibration, Gurugubelli2019LargePeriodicity} to moderate-to-low Reynolds numbers $1000\leq Re \leq 20$ \cite{Ryu2015FlappingFlow,Shoele2016EnergyFlag,Tang2015DynamicsFlow,Gurugubelli2019LargePeriodicity} show that the vibration and vortex dynamics of inverted foils are relatively insensitive to Reynolds number for $Re \geq 200$.
In application to energy harvesting, \cite{kim2013flapping} first proposed an energy harvesting model based on inverted foils, and since then, many have used this configuration for energy harvesting devices \cite{Gurugubelli2015EnergyFoil, Gurugubelli2015Self-inducedFlow,Shoele2016EnergyFlag,Orrego2017HarvestingFlag,Alam2021EnergyFoil}. When compared with the conventional foil, the inverted foil has $O(10^{3})$ times more maximum strain energy than conventional foils for the same parameters \cite{Gurugubelli2015Self-inducedFlow}. In general, the susceptibility towards instability at low flow velocities, ability to undergo LAF in both air and water and superior energy harvesting performance make inverted foils a suitable alternative to conventional foils for harnessing kinetic energy from fluid flows. 

An effective approach to enhance energy harvesting involves employing multiple foils arranged in an array. When two foils are positioned closely, the resulting vortex sheets can induce fluctuations in the fluid forces acting on the foil. Utilizing these unsteady vortex sheets has been shown to boost energy extraction in large-scale systems such as wind and hydrokinetic turbines. \cite{dabiri2011potential} investigated vertical-axis wind turbines in different arrangements and demonstrated that an optimal arrangement can increase the electric power extracted by the wind farm. Correspondingly, in an inverted foil array setup, a similar enhancement of energy harvesting capabilities is expected in inverted foils owing to the increased unsteadiness in flow. 
Several studies have explored the behavior of two or more foils arranged in different array configurations, including tandem, staggered, and side-by-side setups. \cite{Hu2020-tandem} experimentally studied the performance of two inverted foils in tandem setup while  \cite{Huang2018CouplingFlow} performed numerical analyses on the coupled dynamics of two inverted foils in both tandem and staggered arrangements.They observed that beyond a threshold separation distance, the dynamics of the front foil are predominantly unaffected by the rear foil. The front foil behaves akin to a bluff body, generating periodic vortices in its wake which interact with the downstream rear foil. Depending on the separation distance, this wake-body interaction can either enhance or diminish the flapping performance of the rear foil.  

\cite{Huertas-Cerdeira2018CoupledFlags} and \cite{Hu2020} experimentally investigated the dynamics of two side-by-side inverted foils. These studies revealed that in the presence of a neighboring foil, instability is induced in both foils at a higher stiffness as compared to and isolated inverted foil. Proximity effects, depending on separation distance, also lead to multiple coupled flapping modes in the two foils. Furthermore, \cite{Huertas-Cerdeira2018CoupledFlags} observed significant increases of up to $36\%$ in angular flapping amplitude and $13\%$ in flapping frequency in the side-by-side arrangement compared to when the foil is isolated. In a recent study, \cite{Mazharmanesh2020EnergyArrangements} evaluated the maximum energy harvesting performance of two inverted foils in all three arrangements, and for the optimal arrangement observed up to $16\%$ and $12.5\%$ increase in the power extracted by the rear foil as compared to the upstream and isolated foil, respectively. Overall, these studies demonstrate the substantial influence of proximity and wake interference effects on the flapping response of an inverted foil in various array configurations and its effectiveness in harnessing fluid kinetic energy. 

In this work, our intent is to characterize the interference effects on a flexible inverted foil when it is subjected to the unsteady wake flow of an upstream cylindrical bluff body. Bluff bodies submerged in fluid flow have the tendency to undergo flow-induced vibrations for a range of flow velocities \cite{williamson1988vortex,Jaiman_FIV}. Multi-body systems made up of these cylindrical bluff bodies are found in a myriad of civil, mechanical, and nuclear engineering applications. The interference effects between two/multiple cylindrical tandem bluff bodies have been extensively investigated \cite{zdravkovich1987effects,assi2006experimental,assi2010wake,mysa2016origin} with the aim of mitigating flow-induced vibrations and preventing potential failures.   
In the context of flexible structures, the effect of an upstream bluff body on the dynamics of an elastic inverted foil has been investigated in recent studies.  \cite{Allen2001EnergyEel} studied the effect of wake flow on the downstream membrane and demonstrated that vortices generated by an upstream rectangular plate induce periodic deformation in the downstream membrane, leading to a lock-in process, wherein the oscillation frequency of the membrane is synchronized with the cylinder vortex shedding frequency. Likewise, the vortices generated in the cylinder wake can generate different flapping modes in flexible structures. 

Influence of wake flow on the inverted foil studied experimentally \cite{akaydin2010wake, kim2017dynamics,Alben2021CollectiveEfficiency,ojo2022flapping} and numerically by \cite{ojo2022flapping} for different separation distances between the two structures and different bluff body diameters. The results observed in these studies show that along with proximity with the upstream bluff body, the critical velocity at which the flapping modes occur is influenced by cylinder diameter. An additional small amplitude flapping mode was observed in these studies, during which the foil's motion is synchronized to the cylinder vortex shedding. Much akin to the wake-induced vibration dynamics observed for two tandem structures \cite{mysa2016origin}. However, the maximum amplitude and frequency pattern observed during LAF is similar regardless of the foil position \cite{kim2017dynamics}. \cite{akaydin2010wake} also assessed the energy harvesting performance for flexible cantilevered piezoelectric beams when placed behind a cylindrical bluff body and observed that maximum energy is harvested during this lock-in process. 

 Through this literature review, we observe that, in a uniform flow,  the physics behind its large-amplitude flapping response is due to both vortex shedding and the foil's fluid-elastic instability \cite{Gurugubelli2019LargePeriodicity, Tavallaeinejad2020InstabilityFlow}. When placed behind a cylindrical bluff body, the inverted foil is subjected to the perturbed wake flow of the other. As a result of this wake-body interaction, the foil's dynamics become intricate and governed by the proximity of the upstream body. While the impact of the unsteady wake on the inverted foil flapping dynamics is explored with respect to the geometric parameters like separation distance and cylinder diameter, however, the underlying physics leading to a flexible foil's self-excited flapping response when placed in the wake flow is not completely understood and requires further investigation. 

 A complete understanding of these complex dynamics is necessary to design efficient energy harvesters. Our goal is to help answer the following key questions: (i) How does the wake interference influence the foil's flapping response  (ii) what is the intrinsic relationship between the wake vortices and different flapping modes, (iii) What is the optimal range of bending stiffness and mass ratio to sustain the synchronized motion?  In this study, we consider a flexible inverted foil and numerically investigate the effect of an upstream cylindrical bluff body on its dynamics. Through a series of coupled fluid-structure simulations, we investigate the evolution of flapping instability in an inverted foil subject to uniform flow and unsteady wake flow as a function of non-dimensional bending rigidity $K_{B}$ and $m^{*}$ for a given $Re$. 

 The comparison between the isolated and tandem configurations, immersed in uniform flow in the same conditions, has not been studied to the best of the author's knowledge. For a given $m^{*}$ as a function of $K_{B}$, we observe two additional flapping regimes before the onset of the large amplitude flapping regime: small and moderate amplitude flapping. In contrast to the isolated foil, the downstream foil's response in these modes is sensitive through both $m^{*}$ and $K_{B}$. We characterize these dynamics through our simulation results and derive a new non-dimensional parameter that combines the effect of the foil's inertia and elastic forces on the foil's response and provides a unified scaling of the system. We also visualize the evolution of vortex patterns as a function of $K_{B}$ for these flapping response regimes. 

This paper is organized as follows. In Sec \ref{sect:prob}, the problem setup and in Sec \ref{sect:Num}, the governing equations and the coupled fluid-structure numerical framework are described. Sec \ref{sect:Compdomain} presents the computational domain and mesh convergence. Finally, Sec \ref{sec:Res} covers the detailed results and analysis of the inverted foil, namely, the response characteristics, the vortex organization, the effect of mass ratio and proposed scaling relations.

\section{\label{sect:prob} Problem Setup}
\begin{figure}
 \centerline{\includegraphics[width=0.5\textwidth]{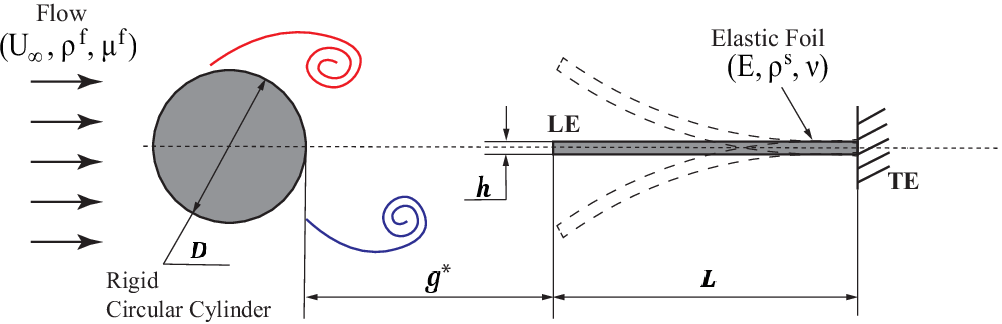}}
  \caption{Schematic of an inverted foil placed downstream of a rigid cylinder}
\label{fig:schem1}
\end{figure}

In our study, we consider a three-dimensional thin, flexible foil $\Omega_{s}$,  which is placed in tandem to a rigid circular cylinder. This setup is subject to uniform incompressible axial flow $\Omega_{f}(t)$, As shown in Fig. \ref{fig:schem1}, the foil is arranged in the \textit{inverted} configuration, wherein its trailing edge (TE) is clamped, and the leading edge (LE) is left free to oscillate. Here, $L$, $W$, and $h$ represent the length, width, and height of the elastic foil, respectively, $D$ represents the diameter of the upstream circular cylinder, and $g$ represents the separation distance between the LE and cylinder surface. $U_{\infty}$ is the magnitude of the uniform free stream velocity, and $\rho^{f}$ and $\mu^{f}$ are the fluid density and dynamic viscosity, respectively. The foil's density is represented by $\rho^{s}$, Young's modulus by $E$, and Poisson's ratio by $\nu$.

The incompressible fluid flow is governed by the Naiver-Stokes equations in the arbitrary Lagrangian-Eulerian (ALE) reference frame, which are defined as:
\begin{equation}
    \rho^{f}\frac{\partial \mathbf{u}^{f}}{\partial t} + \rho^{f}\left(\mathbf{u}^{f} - \mathbf{w}\right)\cdot\nabla\mathbf{u}^{f} =\nabla\cdot\boldsymbol{\sigma}^{f} + \mathbf{f}^{f} \quad \text{in}\; \Omega^{f}(t), \label{eq:Ns1}
\end{equation}
\begin{equation}
\nabla\cdot\mathbf{u}^{f} =0 \quad \text{in}\; \Omega^{f}(t), \label{eq:Ns2}
\end{equation}
\begin{eqnarray}
&&\boldsymbol{\sigma}^{f} = -p\mathbf{I}+\mathbf{T}, \quad \mathbf{T} = 2\mu^{f}\boldsymbol{\epsilon}^{f}(\mathbf{u}^{f}) \nonumber, \\
&&
\boldsymbol{\epsilon}^{f}(\mathbf{u}^{f}) = \frac{1}{2}\left[\nabla\mathbf{u}^{f} + (\nabla\mathbf{u}^{f})^{T} \right], \label{eq:stress}
\end{eqnarray}
where $\mathbf{u}^{f} = \mathbf{u}^{f}(\mathbf{x},t)$ and $\mathbf{w} = \mathbf{w}(\mathbf{x},t)$ represent the fluid and mesh velocities defined at each spatial point $\mathbf{x} \in \Omega^{f}(t)$ respectively.  $\boldsymbol{\sigma}^{f}$ represents the Cauchy stress tensor for a Newtonian fluid as defined in equation (\ref{eq:stress}), and $\mathbf{f}^{f}$ denotes the body force applied on the fluid, and $p$ represents the fluid pressure. $\mathbf{I}$ denotes the second-order identity tensor and $\mathbf{T}$ is the fluid viscous stress tensor.

The elastic foil's deformation is governed by the non-linear structural equation, defined as:
\begin{equation}
 \rho^{s}\frac{\partial \mathbf{u}^{s}}{\partial t}= \nabla\cdot\boldsymbol{\sigma}^{s} + \mathbf{f}^{s} \quad \text{in}\; \Omega^{s}, \label{eq:soliddef}
\end{equation}
where $\mathbf{u}^{s} = \mathbf{u}^{s}(z,t)$ is the structural velocity defined at each point on the structure $\mathbf{z} \in \Omega^{s}$, $\mathbf{f}^{s}$ represents the fluid forces applied on the solid, and $\boldsymbol{\sigma}^{s}$ denotes the first Piola Kirchoff stress tensor. In this study, the structure is modeled as a hyperelastic material using the St. Venant-Kirchoff material. In this coupled fluid-structure system, the velocity and traction continuity conditions are satisfied along the interface as follows:

\begin{eqnarray}
        &&\int_{\boldsymbol{\varphi}^{s}(\gamma,t)} \boldsymbol{\sigma}^{f}(\boldsymbol{\varphi}^{s}(\mathbf{z},t),t)\cdot\mathbf{n}^{f} d\Gamma \nonumber\\
        &&+\int_{\gamma} \boldsymbol{\sigma}^{s}(\mathbf{z},t)\cdot\mathbf{n}^{s} d\Gamma =0\quad \forall \gamma \subset \Gamma_{0}, 
\end{eqnarray}
\begin{equation}
    \mathbf{u}^{f}(\boldsymbol{\varphi}^{s}(z,t),t) = \mathbf{u}^{f}(z,t)\quad \forall \mathbf{z}\in \Gamma_{0}, \label{eq:trac}
\end{equation}
where $\mathbf{n}^{f}$ and $\mathbf{n}^{s}$ are the outward normals to the deformed fluid and the undeformed solid interface boundaries, respectively. $\Gamma_{0}$ denotes the fluid-structure interface $\Gamma(t)$ at time $t = 0$. $\mathbf{\varphi}^{s}$ is the displacement funtion that maps each Lagrangian point $\mathbf{z} \in \Omega^{s}$ to its deformed position at time t. $\gamma$ is any part of the interface $\Gamma_{0}$ and $\mathbf{\varphi}^{s}(\gamma,t)$ represents the corresponding fluid part over the interface $\Gamma(t)$ at $t$.
%
\section{\label{sect:Num}Numerical Framework}
To numerically analyze this complex fluid-structure interaction problem, we employ a higher-order variational method that is based on the combined field explicit interface formulation. This method has proven to be stable for very low structure-to-fluid mass ratios \cite{Liu2014AEffects} and has demonstrated the ability to accurately simulate the dynamics of a thin, flexible foil \cite{Liu2014AEffects} and \cite{Gurugubelli2015EnergyFoil}. In this formulation, the structure's positions and the ALE mesh velocities are decoupled from the other variables (i.e., structural velocity $(\mathbf{u}^{s})$, fluid velocity $(\mathbf{u}^{f})$, and pressure $(p)$) and explicitly calculated at each time step. 
Consider the $n^{th}$ timestep, the position vector of the deformed solid $\boldsymbol{\varphi}^{s}(\mathbf{z},t = n)$, is determined using the second-order Adam-Bashforth method,
\begin{equation}
   \boldsymbol{\varphi}^{s,n} =    \boldsymbol{\varphi}^{s,n-1} + \frac{2\Delta t}{3}\mathbf{u}^{s,n-1} +\frac{\Delta t}{2}\mathbf{u}^{s,n-2}. 
\end{equation}
The motion of the ALE-mesh nodes in the fluid domain $\Omega^{f}(t)$ are modeled as an elastic material in equilibrium,
\begin{eqnarray}
    &&\nabla\cdot\boldsymbol{\sigma}^{m} = 0,\quad\text{with}\\
    &&\boldsymbol{\sigma}^{m} = (1 + \tau^{m})\left[\left(\nabla\boldsymbol{\eta}^{f}+ \left(\nabla\boldsymbol{\eta}^{f}\right)^{T}\right) + \left(\nabla\cdot\boldsymbol{\eta}^{f}\right)\mathbf{I}\right],
\end{eqnarray}
where $\boldsymbol{\sigma}^{m}$ is the stress experienced by the mesh, and $\boldsymbol{\eta}^{f}$ is the displacement vector of the ALE-mesh coordinates that satisfy the boundary conditions
\begin{equation}
    \eta^{f} =\Biggl\{\begin{array}{ccc}
         & \boldsymbol{\varphi}(\mathbf{z},t)-\mathbf{z}\quad & \forall \mathbf{z}\in\Gamma_{0} \\
         & 0 & \text{on}\; \Gamma^{f}(t)\backslash \Gamma(t).
    \end{array}
\end{equation}
The weak form of the Navier-Stokes equations (\ref{eq:Ns1}) and (\ref{eq:Ns2}) can be written as
\begin{equation}
\begin{array}{cc}
     \int_{\Omega^{f}(t)} \rho^{f}\left(\frac{\partial \mathbf{u}^{f}}{\partial t} + \left(\mathbf{u}^{f} - \mathbf{w}\right)\cdot\nabla\mathbf{u}^{f}\right)\cdot \boldsymbol{\phi}^{f} d\Omega + \\
     \int_{\Omega^{f}(t)} \boldsymbol{\sigma}^{f}\colon\nabla\boldsymbol{\phi}^{f} d\Omega 
     =\int_{\Omega^{f}(t)}\mathbf{f}^{f}\cdot\boldsymbol{\phi}^{f} d\Omega \\ 
     + \int_{\Gamma^{f}_{n}(t)}\mathbf{T}^{f}\cdot\boldsymbol{\phi}^{f} d\Gamma +  
        \int_{\Gamma^{f}(t)}\left(\boldsymbol{\sigma}^{f}(\mathbf{x},t)\cdot\boldsymbol{n}^{f}\right)\cdot\boldsymbol{\phi}^{f}(\mathbf{x}) d\Gamma,  
\end{array} \label{eq:weakNS}
\end{equation}
\begin{equation}
         \int_{\Gamma^{f}(t)}q\nabla\cdot\mathbf{u}^{f} d\Omega = 0. \label{eq:Weakns}
\end{equation}
Here, $\boldsymbol{\phi}^{f} \in H^{1}\left(\Omega^{f}(t)\right)$ and $q \in L^{2}\left(\Omega^{f}(f)\right)$ are the smooth test functions in space for the fluid velocity and pressure respectively, and $\Gamma^{f}_{n}(t)$ denotes the fluid Neumann boundary along which $\boldsymbol{\sigma}^{f} (\mathbf{x},t)\cdot\mathbf{n}^{f} = \mathbf{T}^{f}$. Similarly, the weak form of the non-linear structural equation (\ref{eq:soliddef}) is defined as
\begin{equation}
    \begin{array}{cc}
          \int_{\Omega^{s}} \rho^{s}\left(\frac{\partial \mathbf{u}^{s}}{\partial t}\right) \cdot \boldsymbol{\phi}^{s} d\Omega +  \int_{\Omega^{s}} \boldsymbol{\sigma}^{s}\colon\nabla\boldsymbol{\phi}^{s} d\Omega \\ 
          = \int_{\Omega^{s}}\mathbf{f}^{s}\cdot\boldsymbol{\phi}^{s} d\Omega + \int_{\Gamma^{s}_{n}}\mathbf{T}^{s}\cdot\boldsymbol{\phi}^{f} d\Gamma \\
          +  
        \int_{\Gamma_{0}}\left(\boldsymbol{\sigma}^{s}(\mathbf{x},t)\cdot\boldsymbol{n}^{s}\right)\cdot\boldsymbol{\phi}^{s}(\mathbf{z}) d\Gamma,
    \end{array}  \label{eq:weakSolid}
\end{equation}
where $\boldsymbol{\phi}^{s} \in H^{1}\left(\Omega^{s}\right)$ represents the smooth test function of the structural velocity  and $\Gamma^{s}$ denotes the fluid Neumann boundary along which $\boldsymbol{\sigma}^{s} (\mathbf{z},t)\cdot\mathbf{n}^{s} = \mathbf{T}^{s}$. 
The weak form of the traction continuity equation (\ref{eq:trac}) enforced along the fluid-structure interface is defined as
\begin{equation}
 \int_{\Gamma(t)} (\boldsymbol{\sigma}^{f}(\mathbf{x},t)\cdot\mathbf{n}^{f})\cdot\boldsymbol{\phi}^{f} d\Gamma +  \int_{\Gamma_{0}} (\boldsymbol{\sigma}^{s}(\mathbf{z},t)\cdot\mathbf{n}^{s})\cdot\boldsymbol{\phi}^{s} d\Gamma =0. \label{eq:weaktrac}
\end{equation}
It is important to note that in the above equation, $\mathbf{u}^{f}$ is defined on $\Omega^{f}(t)$ and $\mathbf{u}^{s}$  on $\Omega^{s}$. The condition 
\begin{equation}
    \boldsymbol{\phi}^{f}(\boldsymbol{\varphi}^{s}(\mathbf{z},t)) = \boldsymbol{\phi}^{s}(\mathbf{z})\quad \forall\mathbf{z}\in\Gamma_{0},
\end{equation}
is realized by utilizing a conformal mesh along the interface $\Gamma_{0}$. 
The weak form of the coupled CFEI formulation is constructed by enforcing equation (\ref{eq:weaktrac}) to combine equations (\ref{eq:weakNS}-\ref{eq:weakSolid}). For this weak form we must first introduce the finite-dimensional trial solution space $\mathcal{S}$ along with corresponding test function space $\mathcal{V}$. The weak form of the CFEI formulation can be written as find $(\mathbf{u}^{f},p,\mathbf{u}^{s}) \in \mathcal{S}$ such that $\forall(\boldsymbol{\varphi}^{f},q,\boldsymbol{\varphi}^{s})\in\mathcal{V} $
\begin{equation}
    \begin{array}{c}
     \int_{\Omega^{f}(t)} \rho^{f}\left(\frac{\partial \mathbf{u}^{f}}{\partial t} + \left(\mathbf{u}^{f} - \mathbf{w}\right)\cdot\nabla\mathbf{u}^{f}\right)\cdot \boldsymbol{\phi}^{f} d\Omega \\ 
     + \int_{\Omega^{f}(t)} \boldsymbol{\sigma}^{f}\colon\nabla\boldsymbol{\phi}^{f} d\Omega
    -\int_{\Gamma^{f}(t)}q\nabla\cdot\mathbf{u}^{f} d\Omega \\
    + \int_{\Omega^{s}} \rho^{s}\left(\frac{\partial \mathbf{u}^{s}}{\partial t}\right) \cdot \boldsymbol{\phi}^{s} d\Omega +  \int_{\Omega^{s}} \boldsymbol{\sigma}^{s}\colon\nabla\boldsymbol{\phi}^{s} d\Omega \\
    = \int_{\Omega^{f}(t)}\mathbf{f}^{f}\cdot\boldsymbol{\phi}^{f} d\Omega + \int_{\Gamma^{f}_{n}(t)}\mathbf{T}^{f}\cdot\boldsymbol{\phi}^{f} +\int_{\Omega^{s}}\mathbf{f}^{s}\cdot\boldsymbol{\phi}^{s} d\Omega \\
    + \int_{\Gamma^{s}_{n}}\mathbf{T}^{s}\cdot\boldsymbol{\phi}^{f} d\Gamma, \label{eq:combined}
    \end{array}
\end{equation}

From the above equation of the CFEI formulation, it can be observed that the velocity continuity condition is enforced in the functional space, and the traction continuity condition is built in the weak formulation.  Upon discretization of the coupled formulation, the velocity and traction continuity conditions are enforced simultaneously at each time step. 
\section{\label{sect:Compdomain} Computational domain and mesh convergence}
\begin{figure*}
\includegraphics[width=0.8\textwidth]{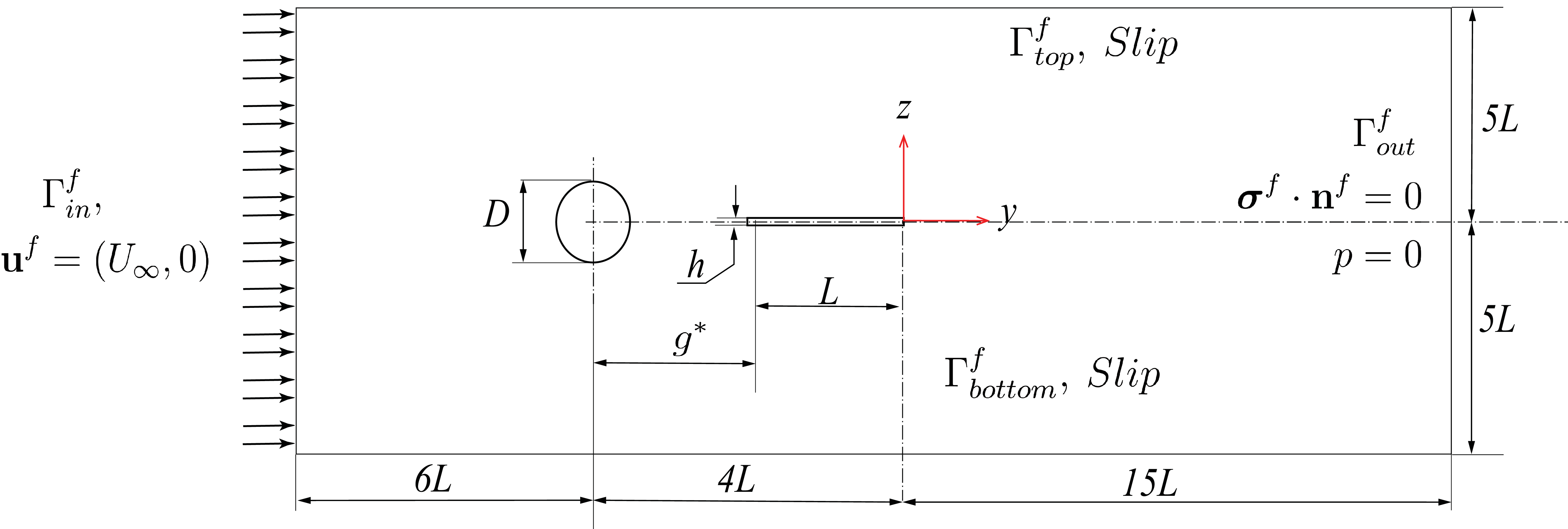}
\caption{Schematic of the computational domain of an inverted foil placed downstream of a stationary circular cylinder}\label{fig:schm}
\end{figure*}
We study the self-induced flapping dynamics of an inverted foil by considering a 2D computational setup. Fig. \ref{fig:schm} displays the computational domain with information on the boundary conditions.  We consider an inverted foil  $\Omega_{s}$ of length $L$, width $W = 0.75L$, and thickness $h= 0.01L$. For the tandem arrangement, the foil is positioned in the downstream region of a rigid circular cylinder of diameter $D$ at a streamwise distance $g^{*}$ along the center line. Both setups are subjected to uniform incompressible axial flow $\Omega_{f}(t)$,  with velocity magnitude $U_\infty$.  The foil is placed with its TE at $(0,0)$. A uniform velocity $\mathbf{u}^{f} = (U_{\infty},0)$ condition is applied at the inlet $\Gamma^{f}_{in}$ boundary, and traction-free condition $(\boldsymbol{\sigma}^{f}\cdot \mathbf{n}^{f}= 0)$ is considered for the outlet $\Gamma^{f}_{out}$ boundary. Here, $\Gamma^{f}_{top}$ and $\Gamma^{f}_{bottom}$ represent the sides of the domain where the slip condition is implemented, and the no-slip condition is applied for the cylinder and foil surfaces. 

A convergence study is performed for three different meshes: $M1$, $M2$, and $M3$ to determine the adequacy of the computation mesh.
The flexible foil in the tandem arrangement has been considered at $Re = 1000$, $m^{*} = 0.1$, and $K_{B} =0.2$. Table \ref{tab:Tab1mesh} lists the r.m.s of the normalized transverse tip displacement ($A^{rms}_{z}/L$) and lift coefficient ($C^{rms}_{l}$), as well as the mean drag coefficient ($\overline{C_{d}}$) for the different meshes.  The bracketed values denote the percentage difference in numerical simulations with respect to mesh $M3$. It is seen that the relative errors using mesh $M2$ are less than $5\%$; therefore, all the numerical simulations are carried out using mesh $M2$. A detailed validation study of the mesh can be found in the Appendix.

\begin{table}
\begin{ruledtabular}
\begin{tabular}{llll}
          & $M1$ &$M2$ & $M3$ \\ \hline
       Nodes   & 36670 & 73194 & 149212\\
       Elements   & 114227 & 223481 & 451083\\
       $A^{rms}_{z}/L$& 0.1039 (59.414$\%$) & 0.245 (4.29$\%$) & 0.256\\
       $\overline{C_{d}}$  & 0.0829 (70.22$\%$) & 0.2864 (2.8735$\%$)& 0.2784\\
       $C^{rms}_{l}$  & 0.5059 (47.32$\%$) & 0.9562 (0.447$\%$) & 0.9605\\

\end{tabular} 
\end{ruledtabular}
\caption{\label{tab:Tab1mesh}Mesh convergence study results for the flexible inverted foil interacting with wake flow of a stationary circular cylinder at $Re = 1000$, $m^{*} = 0.1$, and $K_{B} =0.2$}
\end{table}
 The dynamics of the flexible foil are strongly influenced by three key non-dimensional parameters, i.e., Reynolds number $Re$, bending rigidity $K_{B}$, and mass ratio $m^{*}$  defined as: 
    \begin{equation}\label{1}
        Re=\frac{\rho^{f} U_{\infty} L}{\mu^{f}}, \: K_{B}=\frac{B}{\rho^{f} U_{\infty}^{2} L^{3}}, \: m^{*}=\frac{\rho^{s} h}{\rho^{f} L}.
    \end{equation}
Here $B$ represents the flexural rigidity defined as $B=E h^{3} / 12\left(1-\nu^{2}\right)$ and $K_{B}$ represents the ratio of the foil's bending force to the fluid's inertial force. Where $E$ and $\nu$ denote the Young's modulus and Poisson's ratio of the structure respectively.
We study the derived quantities, including the lift and drag force coefficients. These force coefficients are computed by integrating the element-wise contributions of pressure and viscous stresses for elements located on the foil surface.
The lift coefficient $C_{l}$ and drag coefficient $C_{d}$ are defined as:
\begin{equation}
    C_{l} = \frac{1}{\frac{1}{2}\rho^{f}U^{2}_{\infty}L}\int_{\Gamma}(\rho^{f}\cdot\mathbf{n})\cdot\mathbf{n}_{y} d\Gamma,
\end{equation}
\begin{equation}
    C_{d} = \frac{1}{\frac{1}{2}\rho^{f}U^{2}_{\infty}L}\int_{\Gamma}(\rho^{f}\cdot\mathbf{n})\cdot\mathbf{n}_{z} d\Gamma,
\end{equation}
Here $\mathbf{n}_{y} $ and $\mathbf{n}_{z} $ are the cartesian components of normal vector $\mathbf{n}$.
For a systematic study of the wake body dynamics, we consider a fixed streamwise gap distance $g^{*} = 2.8L$ between the foil and the cylinder and a cylinder diameter of $D/L = 0.4$. This value of $g^{*}$ is greater than the critical distance $4D$ beyond which the co-shedding regime is observed in which the shear layers roll up alternatively to form vortices between the cylinder and the foil \cite{zdravkovich1987effects,sharman2005numerical}. Therefore, the foil only experiences wake interference effects wherein the unsteady wake of the cylinder interacts with the downstream foil and influences the coupled response of the foil. The cylinder dynamics are not influenced by the downstream inverted foil and, thus, are similar to the isolated rigid cylinder case. In this arrangement, the foil's response is investigated at $Re = 1000$ and as a function of $K_{B}$ and $m^{*}$. The Poisson's ratio of the structure is $\nu = 0.3$.

\section{\label{sec:Res} Results and discussion}
\subsection{\label{sub:Ampl}Amplitude response characteristics}
\begin{figure}
 \centerline{\includegraphics[width=0.5\textwidth]{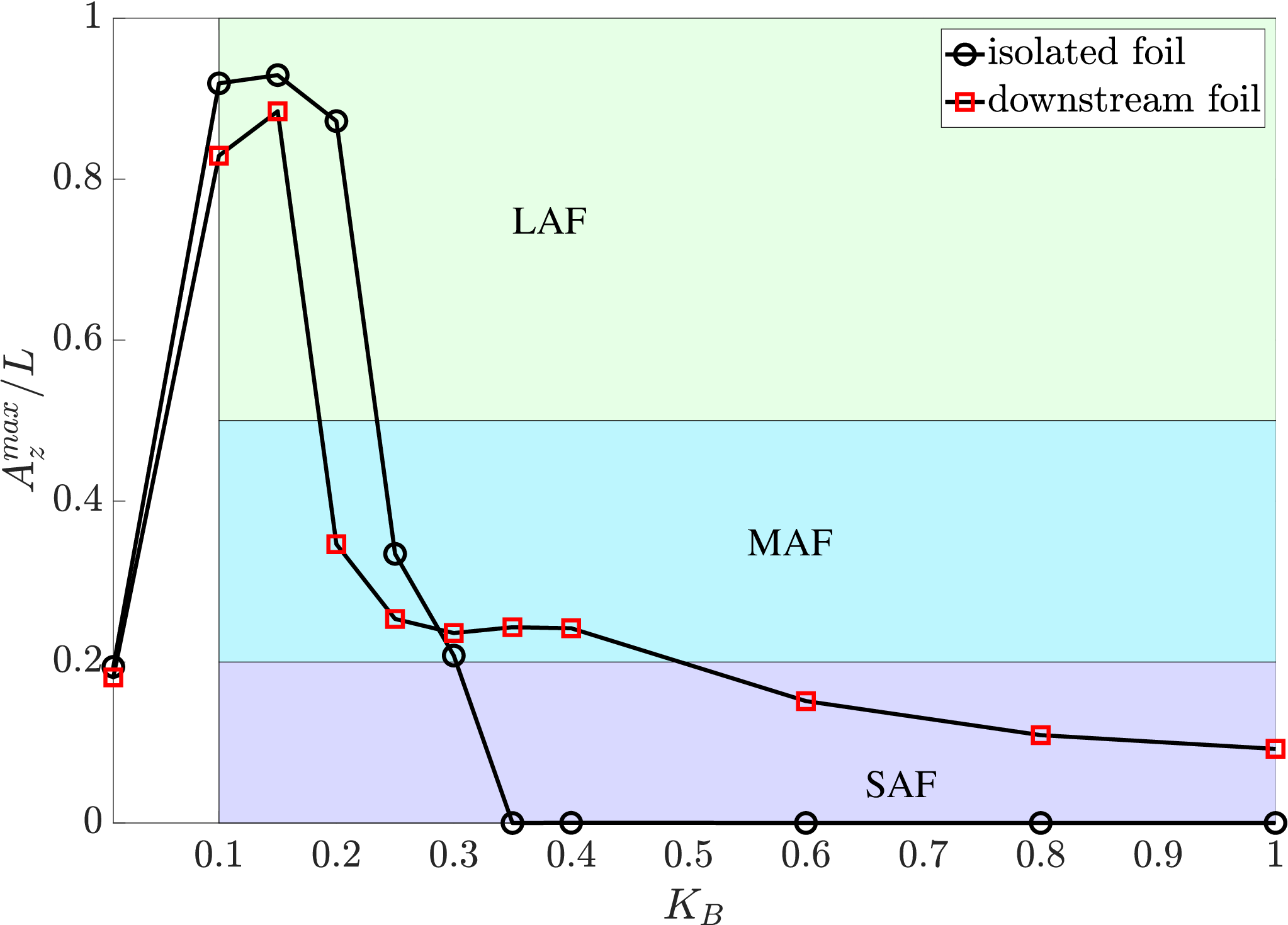}}
\caption{Response characteristics of the inverted foil at $Re = 1000$ and $m^{*} = 0.1$: (a) variation of maximum transverse tip displacement for the isolated and downstream foil as a function of $K_{B}$.}
\label{fig:tip}
\end{figure}

\begin{figure*}
 \centerline{\includegraphics[width=\textwidth]{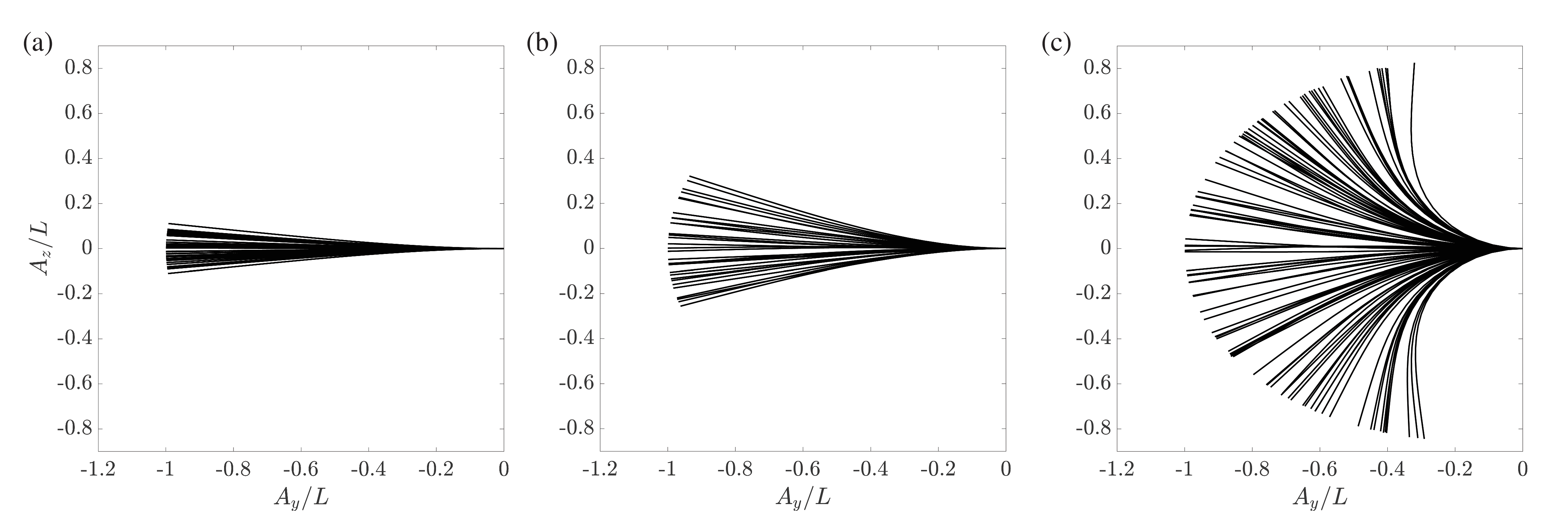}}
  \caption{Full body profiles of the downstream inverted foil at $Re = 1000$ and $m^{*} = 0.1$: (a) small-amplitude flapping, (b) moderate-amplitude flapping, and (c) large-amplitude flapping regimes} 
  \label{fig:profile}
\end{figure*}
We first investigate how the flapping response changes for an inverted foil when placed in tandem with an upstream bluff body compared to the isolated case.  The foil's coupled dynamics are characterized as a function of decreasing $K_{B}$ at $m^{*} = 0.1$ and $Re = 1000$. Fig. \ref{fig:tip} shows the variation of maximum transverse tip displacements of the foil with respect to $K_{B}$.  The maximum transverse tip displacement is defined as $A^{max}_{z}/L = \sqrt{2}A^{rms}_{z}/L$. Here $A^{rms}_{z}/L$,  calculated using $A^{rms}_{z}/L=\left(\sqrt{\frac{1}{n}\Sigma_{t}(A_{z}-\overline{A_{z}}})^{2}\right)/L$, where $A_{z}$  and $\overline{A_{z}}$ represent the transverse tip displacement and the mean transverse displacement,  respectively. 

Consistent with findings from previous experiments done by  \cite{kim2013flapping}, we observe that the isolated inverted foil remains stable when the non-dimensional stiffness ($K_{B}$) is high ($K_{B}>0.35$). Under these conditions, the foil's elastic restoring force exceeds the external fluid forces. 
However, as $K_{B}$ decreases $K_{B} < 0.35$, the foil's elastic restoring moment is no longer sufficient to balance the fluid moment, leading to static divergence instability and deformation under fluid loading \cite{kim2013flapping}. As $K_{B}$ is decreased, the foil experiences flapping instability and starts vibrating about its mean position.  The maximum transverse oscillation amplitude of the foil tip increases with a decrease in foil stiffness as shown in Fig. \ref{fig:tip}. These transverse vibrations eventually develop into large-amplitude periodic oscillations with $A^{max}_{z}/L > 0.8$ that are observed for $0.1\leq K_{B} \leq 0.2$. The foil transitions from the large-amplitude flapping (LAF) regime to the deflected flapping and eventually flipped flapping regime as $K_{B}$ is further decreased.

In contrast to the isolated foil,  in the tandem arrangement, the inverted foil performs sustained oscillations for $K_{B} \leq 1$. As a result, the range of $K_{B}$ where flapping instability is observed is wider for the inverted foil in the tandem arrangement than an isolated foil. These observations are consistent with previous experimental and numerical studies \cite{akaydin2010wake,ojo2022flapping}.  
Two additional flapping regimes are identified before the downstream foil transitions into the LAF regime based on the foil's maximum transverse tip displacement, namely the (i) small amplitude flapping regime (SAF) when $A^{max}{z}/L < 0.2$ and (ii) moderate amplitude flapping regime (MAF) when $0.2 < A^{max}{z}/L \leq 0.5$. The SAF and MAF regimes are observed for $0.4\leq K_{B}\leq 1$ and $0.2\leq K_{B}<0.4$, respectively. The maximum transverse tip displacement for both regimes gradually increases as $K_{B}$ decreases, resulting in a smoother transition to the LAF regime compared to the isolated case. For $K_{B}<0.2$ the downstream foil transitions into the large amplitude flapping regime undergoing oscillations with transverse amplitudes $A^{max}{z}/L > 0.5$. The LAF regime is exhibited by the downstream foil for a narrower range of $K_{B} \in [0.1,0.15]$ as compared to the isolated case of $K_{B}\in [0.1, 0.2]$. Typical full-body response profiles of the downstream inverted foil in SAF, MAF, and LAF regimes is shown in Fig. \ref{fig:profile}. For $K_{B}<0.1$, as $K_{B}$ is decreased, downstream foil transitions to deflected flapping mode and finally flipped flapping mode for very small $K_{B} \in [0.01,0.001]$ similar to the isolated foil.  In order to understand the distinct dynamics of the inverted foil in the tandem setup and the origin of the two additional flapping modes, for the remaining part of the paper, we will consider the response for an inverted foil in both configurations for $K_{B}\in [0.1, 1]$.

\subsection{\label{sub:freq}Frequency response and force analysis}
\begin{figure}
 \centerline{\includegraphics[width=0.5\textwidth]{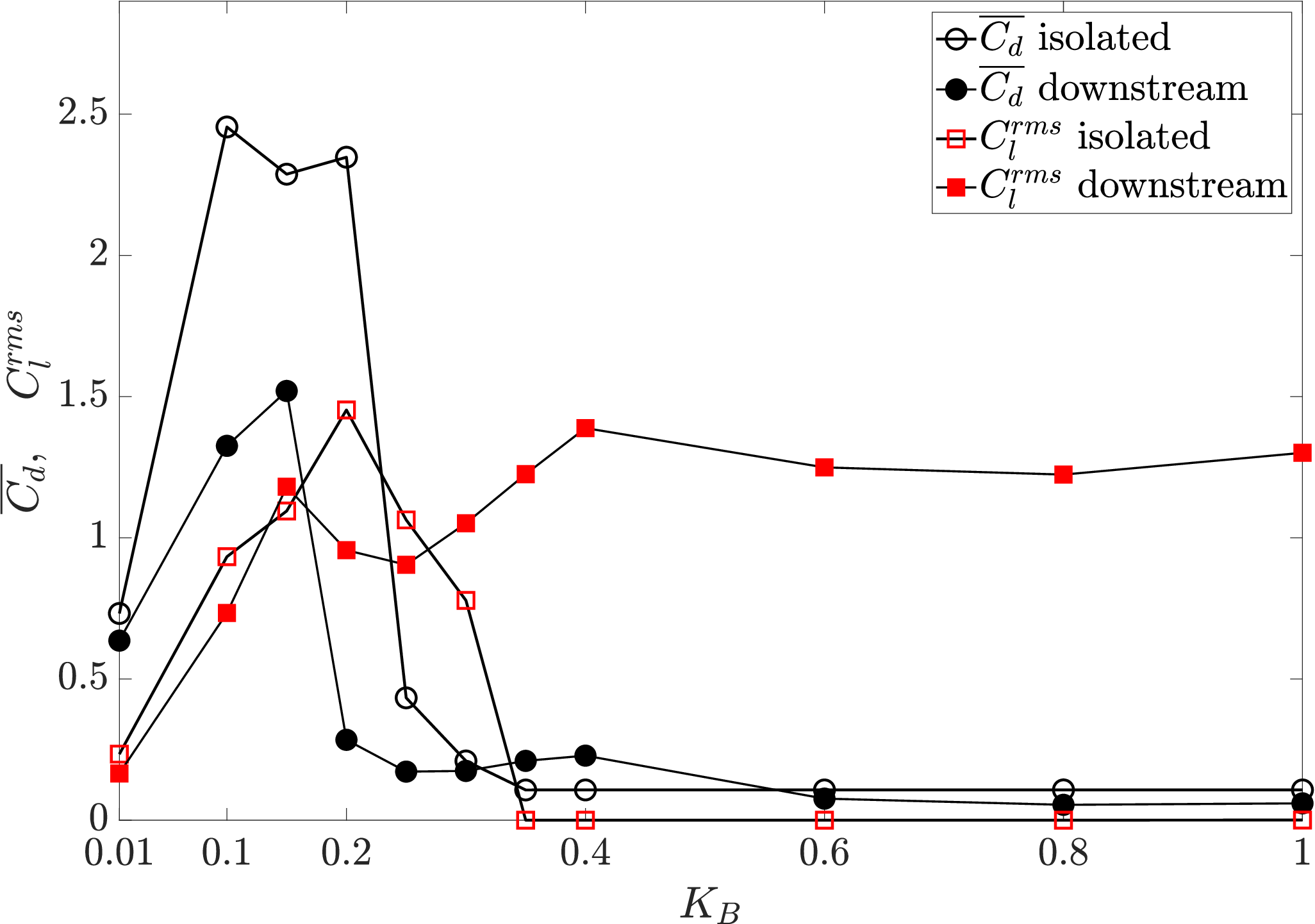}}
  \caption{Variation of fluid loading for the isolated and downstream foil as a function of $K_{B}$}
\label{fig:4}
\end{figure}
 We next investigate the variation in the fluid loading and the frequency characteristics of the inverted foil response when it interacts with the wake flow for $K_{B}\in [0.1, 1]$ at $Re = 1000$ and $m^{*} = 0.1$.  The lift force and transverse displacement frequencies are calculated using the Fast Fourier Transform (FFT) and non-dimensionalized using $L/U_{\infty}$. One must note that at the considered separation distance $g^{*} = 2.8L$, the downstream inverted foil does not influence the cylinder dynamics. Therefore, a constant vortex shedding frequency $f_{vs} = 1.9342$ Hz is observed for all values of $K_{B}$ based on the Strouhal number of $0.21$. 
 Fig.  \ref{fig:4} shows the variation of the mean drag coefficient $\overline{C_{d}}$  and the root mean square of the lift coefficient $C^{rms}_{l}$ with respect to  $K_{B}$ for the inverted foil in both setups. Qualitatively, the drag force exerted on the inverted foil in both setups follows similar trends as a function of $K_{B}$. Wherein the drag force increases as $K_{B}$ decreases, instability is induced, and deformation increases.  In contrast, a net higher lift force is exerted on the downstream foil than in the isolated case. For $K_{B}> 0.35$, a lift coefficient for the downstream foil is around $O(10)$ times higher than the drag force and up to $O(10^3)$ times greater than the lift coefficient for the isolated foil thereby resulting in the early onset of flapping instability. These results demonstrate that for $K_{B}> 0.35$, the oscillations of the downstream foil in the tandem setup are driven by the fluid loading due to the unsteady wake. We attribute this higher fluid loading to the alternating wake vortices shed from the upstream bluff body, which flows above and below the downstream inverted foil \cite{akaydin2010wake}. When a vortex impinges the foil, its low-pressure core creates a low-pressure region on the foil surface. Simultaneously, the induced flow of another wake vortex creates a high-pressure region on the opposite side of the foil. A combined effect of these two pressure mechanisms causes a higher fluid loading on the downstream inverted foil than in the isolated case. The higher fluid loading on the downstream foil results in an early onset of flapping instability, as shown in Fig. \ref{fig:tip}.        

\begin{figure}
 \centerline{\includegraphics[width=0.5\textwidth]{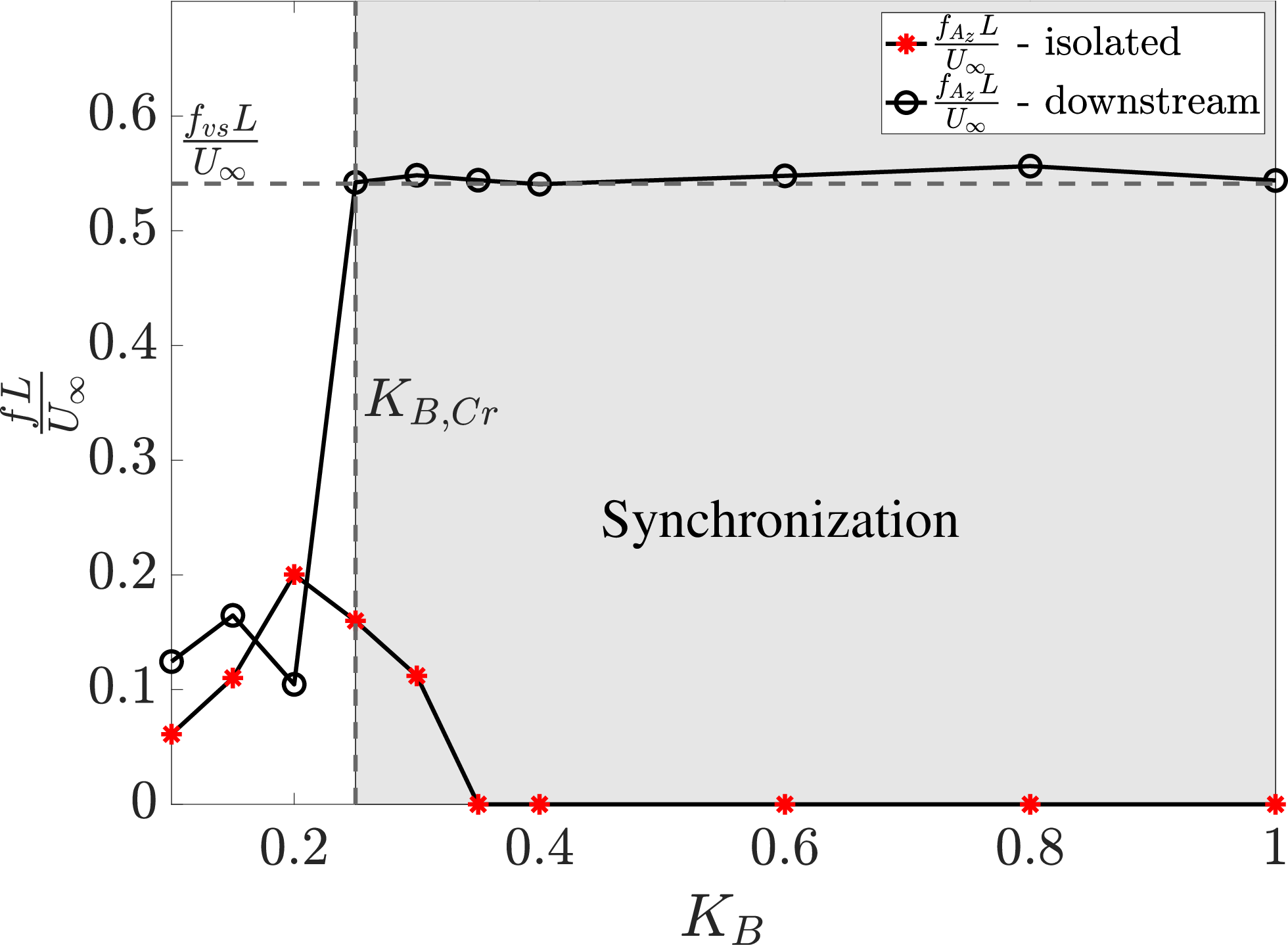}}
  \caption{Variation of the transverse flapping frequencies of the isolated and downstream foil for $K_{B} = [0.1, 1]$, at $Re =1000$ and $m^* = 0.1$}
\label{fig:6}
\end{figure}
\begin{figure}
 \centerline{\includegraphics[width=0.5\textwidth]{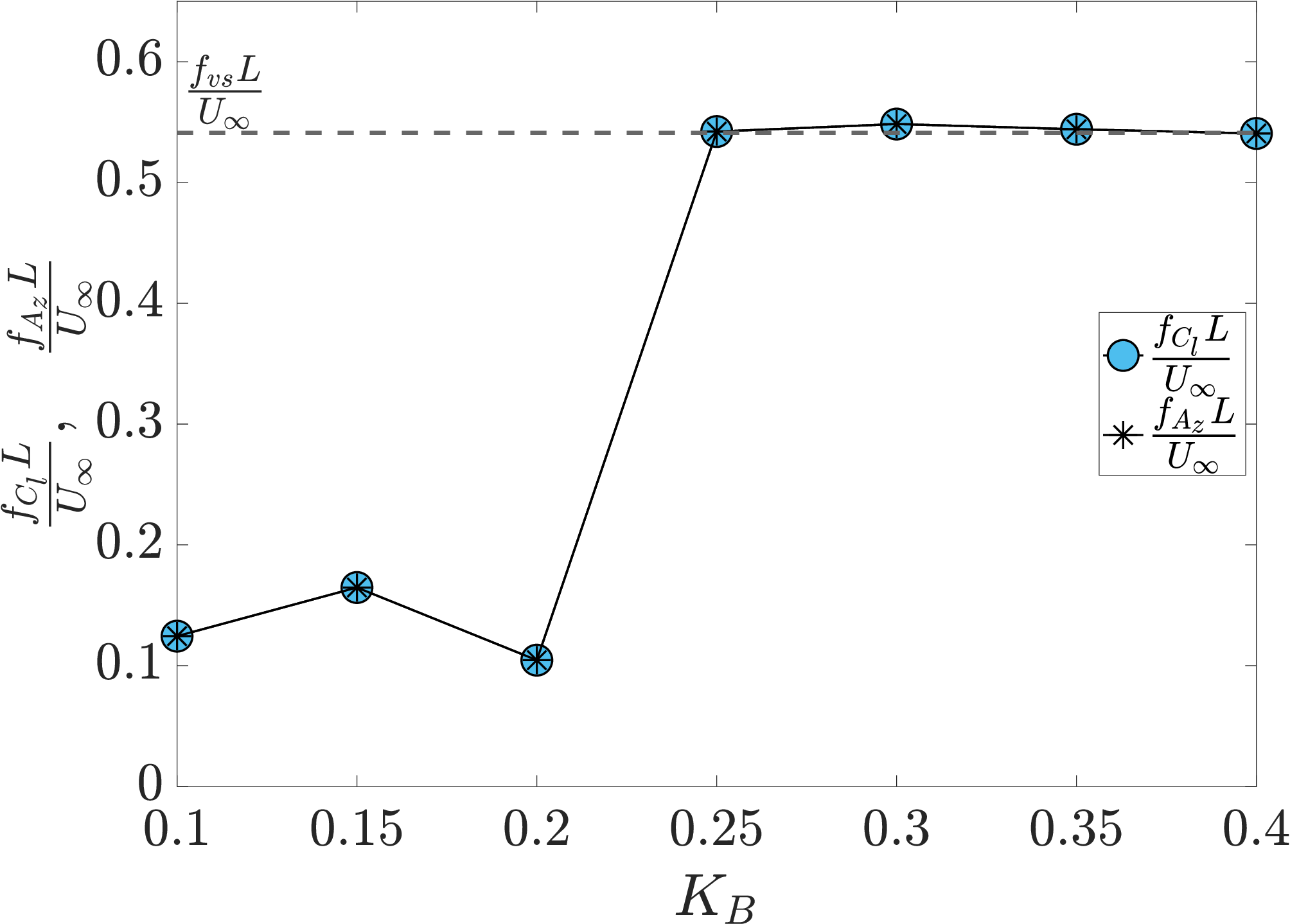}}
  \caption{Dependence of the transverse flapping and lift frequencies on bending stiffness $K_{B}$}
\label{fig:7}
\end{figure}

 Fig. \ref{fig:6}  shows the variation in the dominant non-dimensional transverse flapping frequency $f_{A_{z}}L/U_{\infty}$ of an inverted foil in the isolated and tandem setups as a function of $K_{B}$ at $Re =1000$ and $m^{*} = 0.1$. Fig. \ref{fig:7} demonstrates the variation of the dominant frequency of the transverse force $f_{C_{l}}L/U_{\infty}$ and tip displacement $f_{A_{z}}L/U_{\infty}$ of the downstream foil with $K_{B}$. If we consider the isolated case as shown in Fig. \ref{fig:6}, we observe that once instability is induced at $K_{B} < 0.35$, the inverted foil undergoes low-frequency oscillations as the foil transitions into the large amplitude flapping regime. Once the foil transitions to the LAF regime, the flapping frequency reduces as $K_{B}$ is reduced. Owing to increased flexibility and the resulting increase in flapping amplitude, the time taken for the foil to complete the oscillations increases \cite{kim2013flapping, Gurugubelli2015Self-inducedFlow}.

 In contrast to the isolated foil, the frequency response of the downstream foil demonstrates distinct characteristics as displayed in Fig. \ref{fig:6}. 
 We observe that for $K_{B} \geq 0.25$, the dominant frequency $ f_{A_{z}}\approx f_{vs}$  and is unaffected by the change in foil flexibility. A sharp decline in the dominant frequency ratio is observed for $K_{B} < 0.25$ with $f_{A_{z}}\approx 0.2f_{vs}$.  In our study, a frequency match is also observed between the transverse tip displacement and lift force for all values of $K_{B}$ as shown in Fig. \ref{fig:7}.  Based on these results, we can characterize the response of the downstream inverted foil in two regimes. The first regime, known as the 'lock-in' regime, is observed up to a threshold value of non-dimensional bending rigidity $K_{B} = 0.25$, wherein the cylinder wake strongly influences the inverted foil's dynamics, and the foil's motion is synchronized with the cylinder vortex shedding frequency. During synchronization, depending on foil flexibility, the inverted foil undergoes high-frequency oscillations with either small or moderate amplitudes. Below $K_{B} = 0.25$, denotes the post 'lock-in' regime.  A similar synchronization phenomenon is observed in experiments conducted by \cite{akaydin2010wake} and \cite{ojo2022flapping} in which the inverted foil dynamics are controlled by the cylinder vortex shedding.  In this regime,  the influence of the wake reduces, and the foil is no longer synchronized with the cylinder wake. As a result, when  $K_{B}$ is reduced, the downstream foil rapidly oscillates in LAF mode with low oscillation frequencies similar to an isolated foil.

\subsection{\label{sub:vorti}Vorticity Dynamics}
\begin{figure*}
         \centering
         \includegraphics[width=\textwidth]{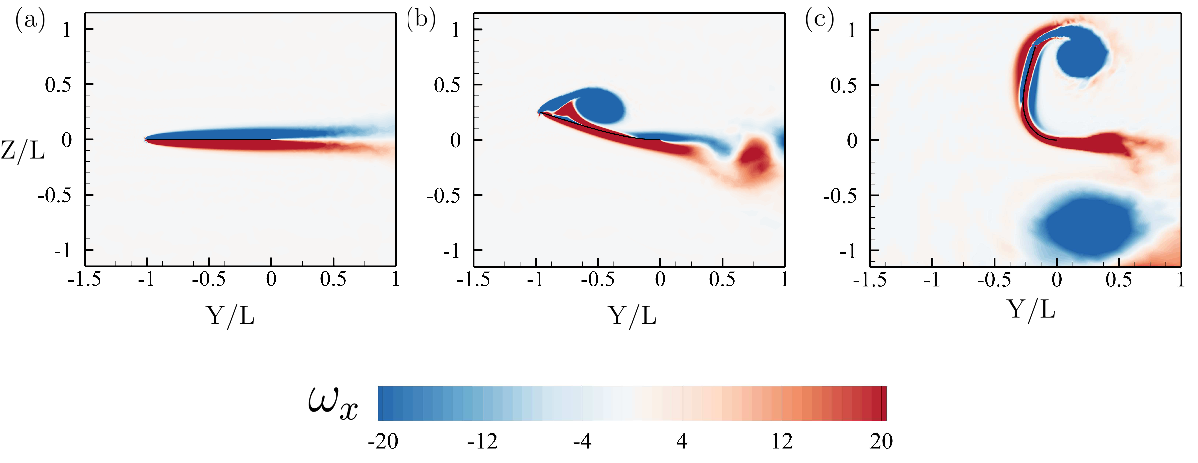}
         \caption{Streamwise vorticity contours for the isolated inverted foil given $Re = 1000$, $m^{*} = 0.1$ at: (a) $K_{B} = 1$, (b) $K_{B} = 0.3$, and (c) $K_{B} = 0.15$    }
         \label{fig:vortiso}
\end{figure*}
\begin{figure*}
         \centering
         \includegraphics[width=\textwidth]{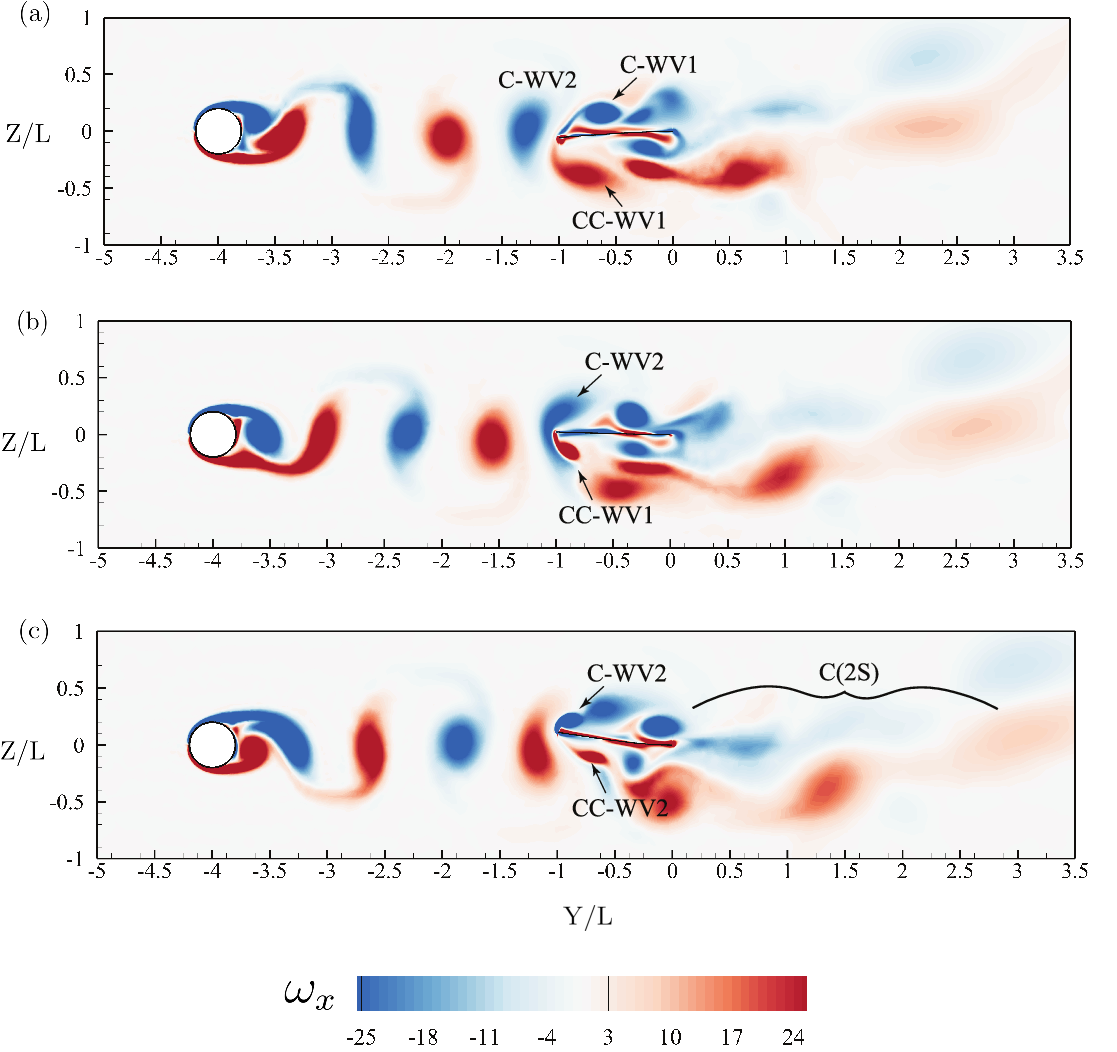}
         \caption{Streamwise vorticity contours for the downstream inverted foil at $K_{B} = 1$, given $Re = 1000$, $m^{*} = 0.1$: (a) $tU/L = 43.75$, (b) $tU/L = 44.28$, and (c) $tU/L = 44.82$  }
         \label{fig:KB1min}
\end{figure*}
\begin{figure*}
         \centering
         \includegraphics[width=\textwidth]{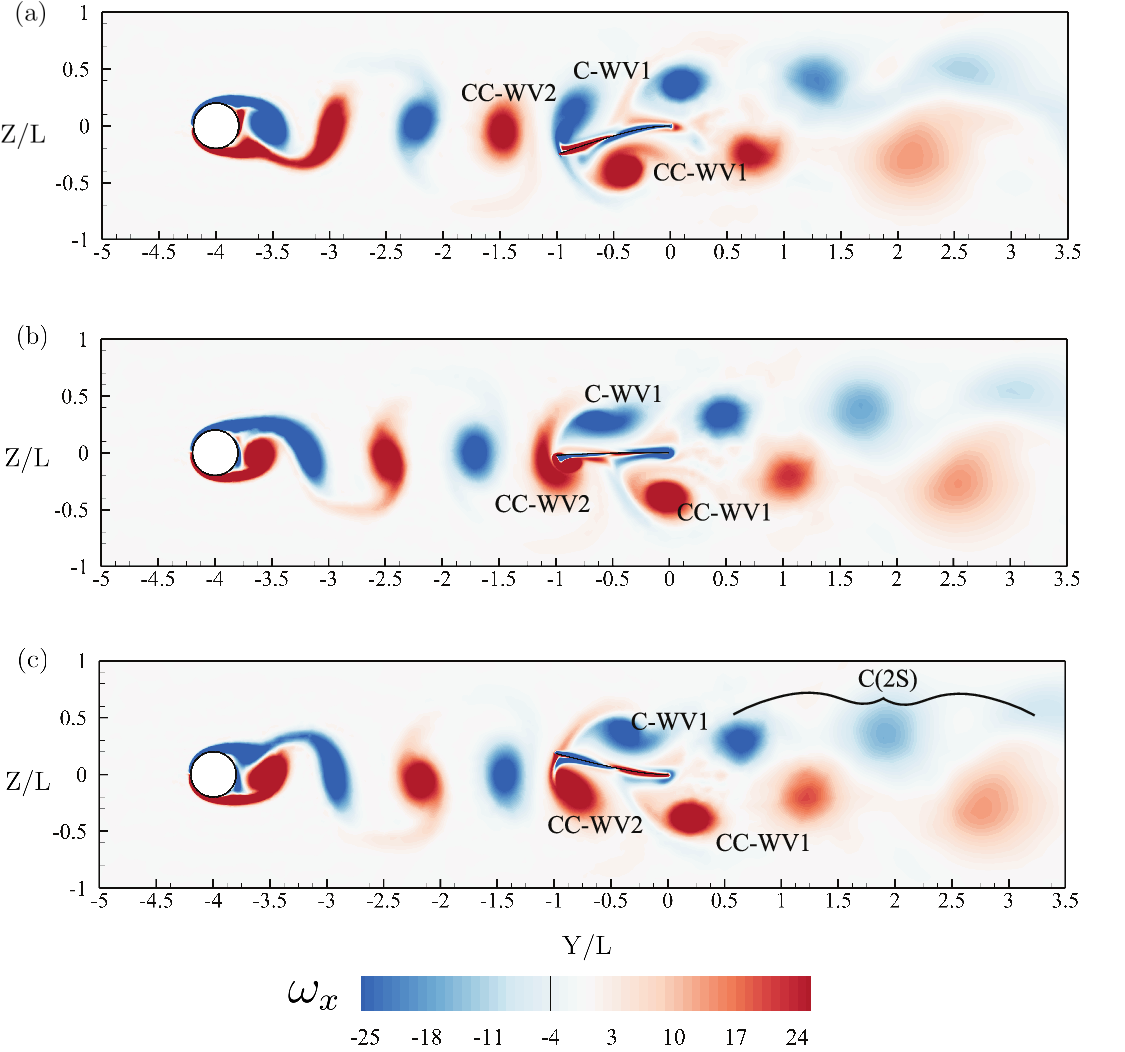}
         \caption{Streamwise vorticity contours for the downstream inverted foil at $K_{B} = 0.3$, given $Re = 1000$, $m^{*} = 0.1$: (a) $tU/L = 44.46$, (b) $tU/L = 45.08$, and (c) $tU/L = 45.446$ }
         \label{fig:KB03cyl}
\end{figure*}
\begin{figure*}
         \centering
         \includegraphics[width=0.9\textwidth]{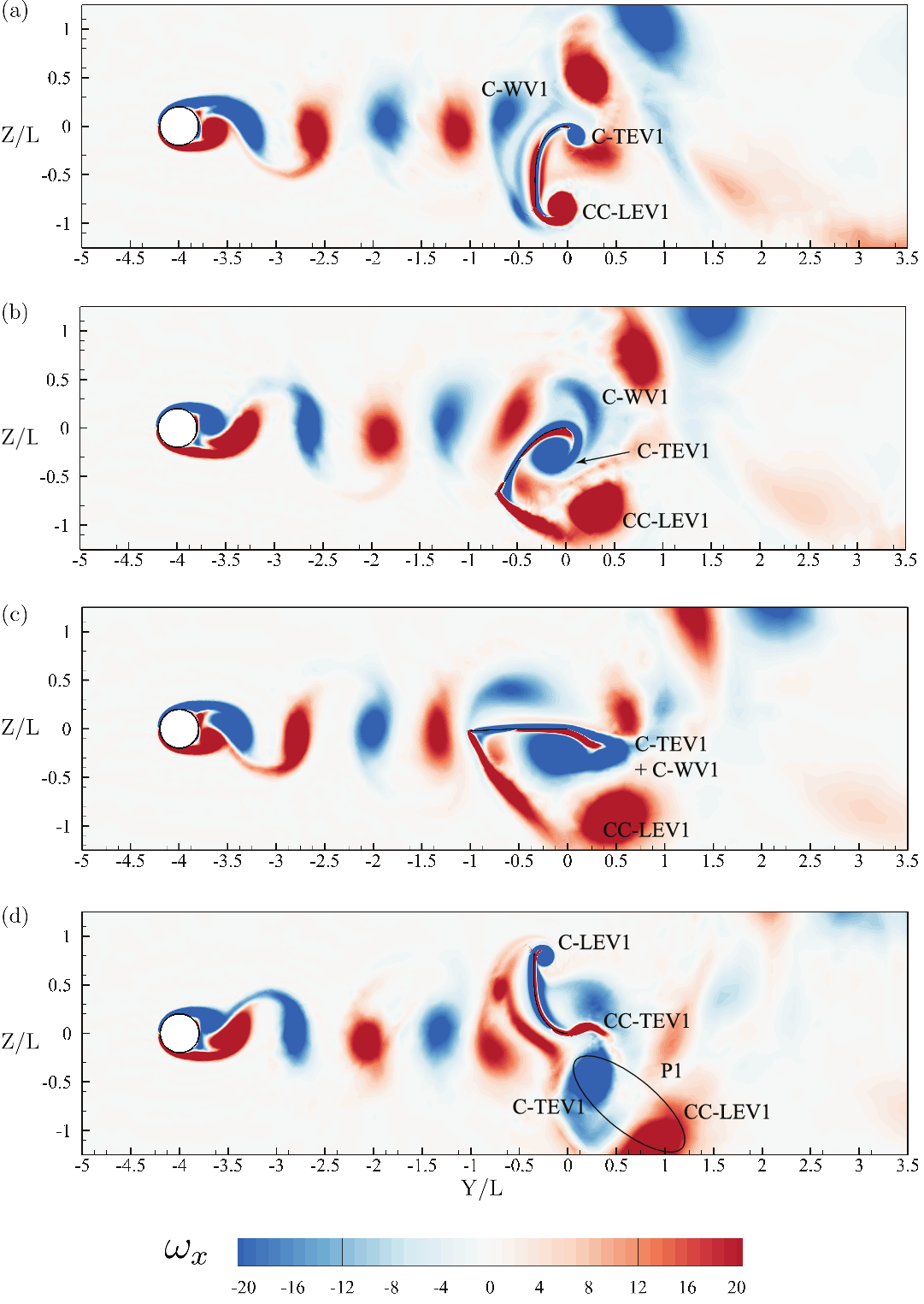}
         \caption{Streamwise vorticity contours for the downstream inverted foil at $K_{B} = 0.15$, given $Re = 1000$, $m^{*} = 0.1$: (a) $tU/L = 41.78$, (b) $tU/L = 42.5$, (c) $tU/L = 43.48$, and (d) $tU/L = 44.55$  }
         \label{fig:KB015cyl}
\end{figure*}

In this section, we present instantaneous vorticity fields for the inverted foil in both configurations, offering insights into the underlying mechanisms driving the dynamic responses elucidated in Sec \ref{sub:Ampl} and \ref{sub:freq}. Fig. \ref{fig:vortiso}a showcases the streamwise vorticity contours for the isolated inverted foil at $K_{B} = [1, 0.3, 0.15]$, while Figs. \ref{fig:vortiso}(b-c) illustrate the vorticity contours for the foil at the maximum tip displacement above $\mathrm{Z/L} = 0$ during upstroke at $K_{B} = 0.3 \text{ and } 0.15$, respectively. Figs. \ref{fig:KB1min} -\ref{fig:KB015cyl} depict the streamwise vorticity contours for the downstream inverted foil during one half-cycle (upstroke) at $K_{B} = 1$, $K_{B} = 0.3$, and $K_{B} = 0.15$, corresponding to the small amplitude flapping (SAF), moderate amplitude flapping (MAF), and large amplitude flapping (LAF) regimes, respectively.

At $K_{B} =1$, the isolated inverted foil is in the fixed point stable regime, and the flow field typically represents boundary layer flow over a flat plate as shown in Fig. \ref{fig:vortiso}a. Symmetric boundary layer flow is developed on either side of the inverted foil as fluid travels across the foil. These shear layers form a narrow, steady wake as the flow travels off the fixed trailing edge.  In the tandem arrangement, as fluid travels over the cylinder, a steady Karman wake is generated in the cylinder wake made up of two oppositely signed single vortices convecting downstream with a uniform transverse displacement. As described by \cite{williamson1988vortex}, this is known as the 2S shedding mode. The wake flow consisting of high-energy vortices shed from the cylinder interacts with the boundary layer of the downstream inverted foil, leading to significantly higher loading experienced by the downstream foil, as observed in Sec \ref{sub:freq}.

Fig. \ref{fig:KB1min} shows the vortex organization of downstream foil over one upstroke cycle at $K_{B} = 1$.  
As shown in Fig. \ref{fig:KB1min}a,  when the foil is at the maximum deformation below $\mathrm{Z/L} =0$, i.e., at the beginning of upstroke, a (negative signed) clockwise wake vortex (C-WV1) interacts with the same signed shear layer formed on the upper surface of the foil at the leading edge creating a low-pressure region on the upper surface of the foil.  Meanwhile, a positive signed counterclockwise wake vortex (CC-WV1) reaches the foil tip and flows below the foil, and the low-velocity fluid surrounding the vortex creates a high-pressure region on the lower surface of the foil. As a result, the foil starts moving in an upward direction.  As the foil is about to cross the line $\mathrm{Z/L} = 0$, C-WV1 travels along the foil into the wake, while CC-WV1 merges with the same signed shear layer below the foil and vortices in the wake below the foil that builds up the high-pressure region below the foil. Simultaneously clockwise wake vortex (C-WV2) impinges the foil at the leading edge and combines with the same-signed shear layer on the upper surface of the foil. Under the combined effect of wake vortices flowing in the wake and interacting with the shear layer at the leading edge, the foil reaches its maximum deformation above $\mathrm{Z/L} = 0$. The same process is repeated during the downstroke, except all the vortices are reversed. If we consider the wake topology, a total of two single opposite signed vortices (2S) are shed from the foil per one time period that convects into the wake, similar to the stationary cylinder. 

As illustrated in Fig. \ref{fig:KB1min}c, vortices shed during the current cycle coalesce with the same signed vortices shed during the previous cycles, resulting in a C(2S) vortex shedding pattern observed in the near wake of foil as described by \cite{williamson1988vortex}. Typically, in the SAF regime, the downstream foil undergoes sustained small amplitude oscillations induced by the interaction between the alternating vortices shed by the cylinder and the shear layer developed on either side of the foil as the fluid travels across the foil. Consequently, the foil oscillates at a frequency equal to the vortex shedding frequency of the upstream bluff body, as shown in Fig. \ref{fig:tip}. 

As the non-dimensional frequency of the foil is reduced $K_{B} < 0.4$, the downstream foil transitions to the MAF regime, as discussed in Sec \ref{sub:Ampl}.  Fig. \ref{fig:KB03cyl}a demonstrates that similar to the wake topology observed for the downstream foil during SAF mode, in the MAF mode, at the beginning of upstroke, the positive signed CC-WV1 flows below the foil, while negative signed (C-WV1) vortex impinges the foil at the leading edge of the foil. However, the downstream foil oscillates with larger flapping amplitudes due to increased flexibility at these $K_{B}$ values. Consequently, a leading edge vortex (LEV) is formed at the leading edge, which merges with the same signed wake vortex C-WV1. This merging results in the separation of the leading edge vortex that flows into the wake as the foil approaches $\mathrm{Z/L} = 0$. As shown in  Fig. \ref{fig:KB03cyl}b, a counter-rotating leading edge vortex is formed at the leading on the lower surface of the foil that merges with an approaching counter-clockwise wake vortex (CC-WV2) when the foil crosses $\mathrm{Z/L} = 0$. This combined LEV and CC-WV2 grow as it reaches its peak position during the upstroke, as illustrated in Fig. \ref{fig:KB03cyl}c. The separation of combined-(CC-WV2) marks the end of the upstroke and the onset of the downstroke.

Generally, two opposite signed vortices are shed into the near wake of the foil during one oscillation cycle for the MAF regime similar to the SAF. However, due to larger deformation, two layers of the same signed vortices are formed above and below the foil. These vortices coalesce together to form the C(2S) pattern in the near wake of the foil, as shown in Fig. \ref{fig:KB03cyl}. The downstream foil experiences the MAF regime for $0.2\leq K_{B}<0.4$. For these values of $K_{B}$, flapping instability is induced in the inverted foil even when it is submerged in uniform flow, as demonstrated in Fig. \ref{fig:vortiso}a. As a result, a change in wake-body interaction is observed in this regime for the downstream foil compared to the SAF regime. These results suggest that, unlike the SAF regime in the MAF regime, the wake vortices control the flapping amplitude and frequency and prevent transition to the LAF regime instead of inducing the flapping motion.

For $K_{B} < 0.2$, the downstream foil experiences large amplitude oscillations similar to the isolated foil.  Fig. \ref{fig:KB015cyl}a represents the instantaneous vorticity corresponding to the condition when the leading edge is at the lowest position, a pair of oppositely signed vortices: counterclockwise leading edge vortex (CC-LEV1) and clockwise trailing edge vortex (C-TEV1) are formed at the leading and trailing edges of the foil. As observed in Fig. \ref{fig:KB015cyl}b, the CC-LEV1 develops and pairs with the negatively signed C-TEV1 as the foil approaches $\mathrm{Z/L} = 0$. Due to the large deformation, the wake vortices flow above the foil with minimal interaction with the foil. The pair of vortices CC-LEV1 + C-TEV1 grow and separate as the foil crosses the centerline $\mathrm{Z/L} = 0$, and are eventually shed into the wake at the end of upstroke as shown in Fig. \ref{fig:KB015cyl}(d). Therefore, two pairs of vortices are shed over one cycle, resulting in a 2P vortex pattern observed in the wake. Similar to the downstream foil, a pair of vortices are also shed from the foil's leading and trailing edge for the isolated foil at $K_{B} = 0.15$ as it completes one downstroke cycle as demonstrated in Fig. \ref{fig:vortiso}c, which leads to a 2P vortex shedding mode. These contour plots show that unlike the SAF and MAF regimes, during LAF, the wake vortices have negligible influence on the response of the foil. 

Through these vorticity contour plots, we observe primarily 2S and 2P vortex-shedding patterns in the wake of the downstream foil depending on the non-dimensional bending rigidity $K_{B}$. The 2S mode is observed for $K_{B} \geq 0.25$, below which 2P mode is observed in the MAF and LAF regimes. Thus, we can deduce that when the inverted foil is synchronized with the vortex shedding of the upstream cylinder, it undergoes 2S vortex shedding like the cylinder. For $K_{B}\leq 0.25$, the foil starts behaving similarly to an isolated inverted foil and no longer oscillates in sync with the cylinder vortex shedding. The consequent change in the flapping frequency results in a transition in the vortex shedding pattern from 2S to 2P mode.

\subsection{ \label{sub:Mass}Effect of mass ratio}
\begin{figure}
 \centerline{\includegraphics[width=0.5\textwidth]{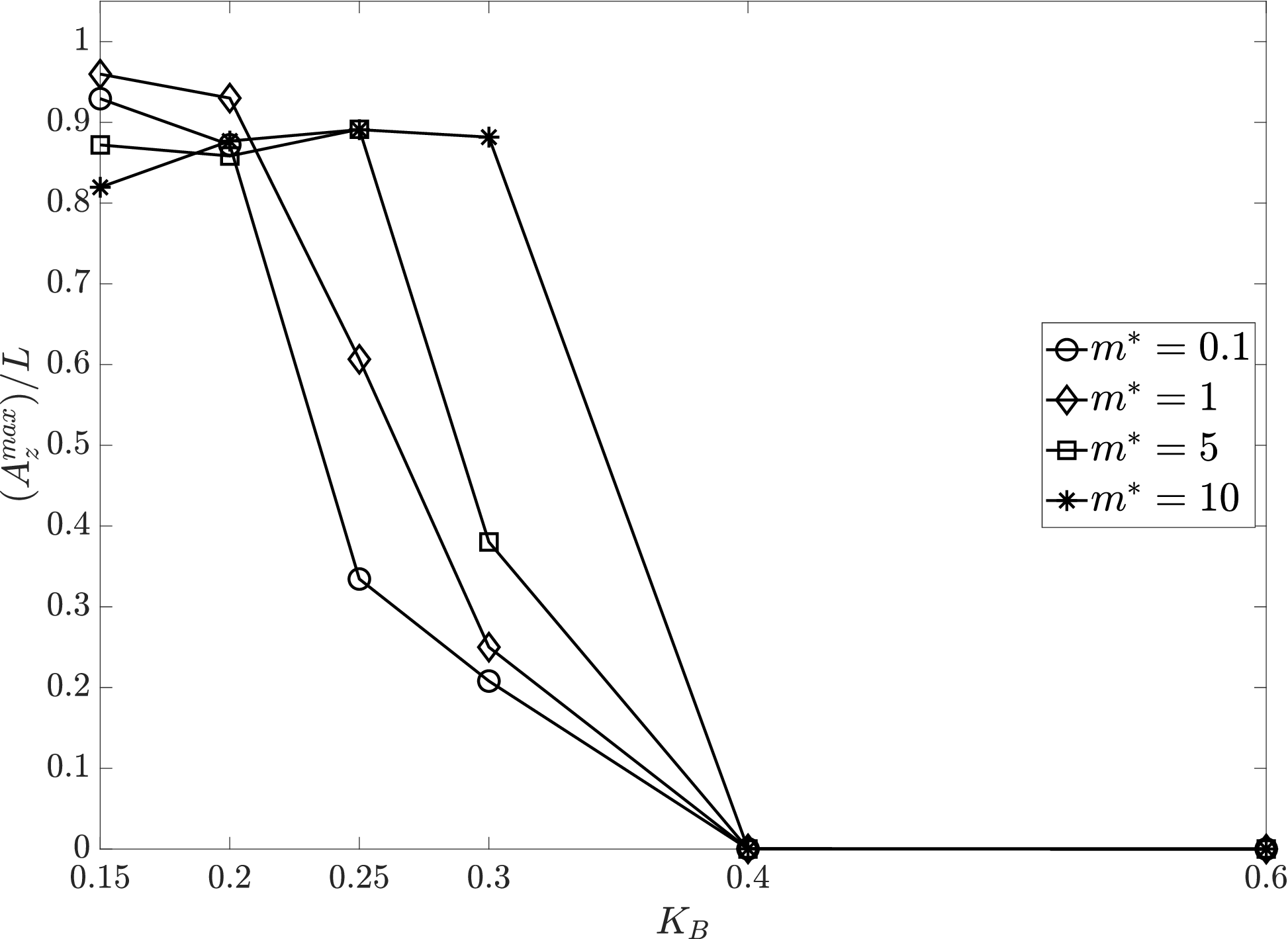}}
  \caption{Variation of maximum transverse tip displacement with non-dimensional bending stiffness $K_{B}\in[0.15, 1]$ and $m^{*}=[0.1,1,5,10]$ for an isolated inverted foil at $Re = 1000$}
\label{fig:8}

\end{figure}
\begin{figure}
 \centerline{\includegraphics[width=0.5\textwidth]{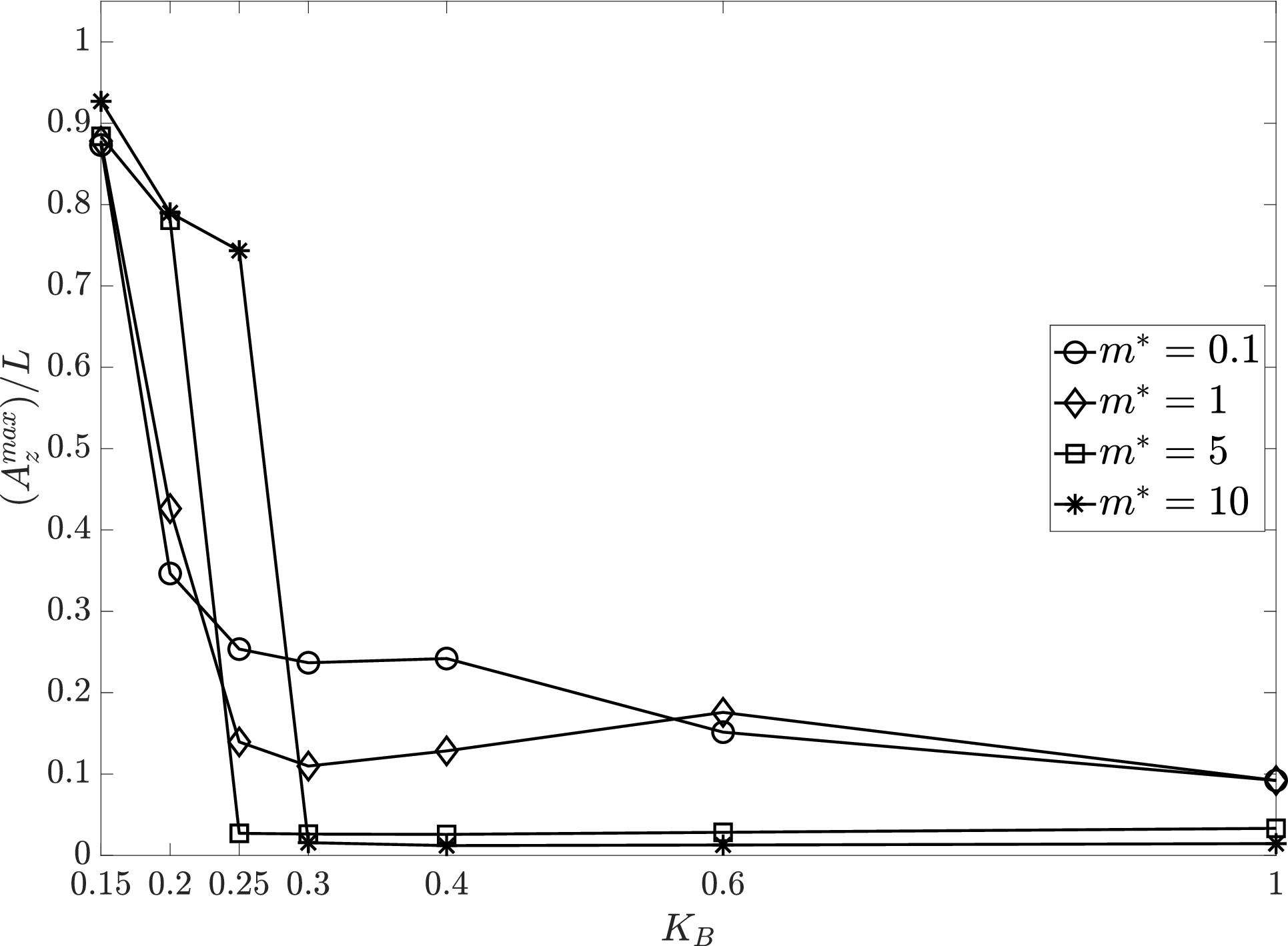}}
  \caption{Variation of maximum transverse tip displacement with non-dimensional bending stiffness $K_{B}\in[0.15, 1]$ and $m^{*} =[0.1,1,5,10]$ for the downstream inverted foil at $Re = 1000$}
\label{fig:9}
\end{figure}

\begin{figure}
 \centerline{\includegraphics[width=0.5\textwidth]{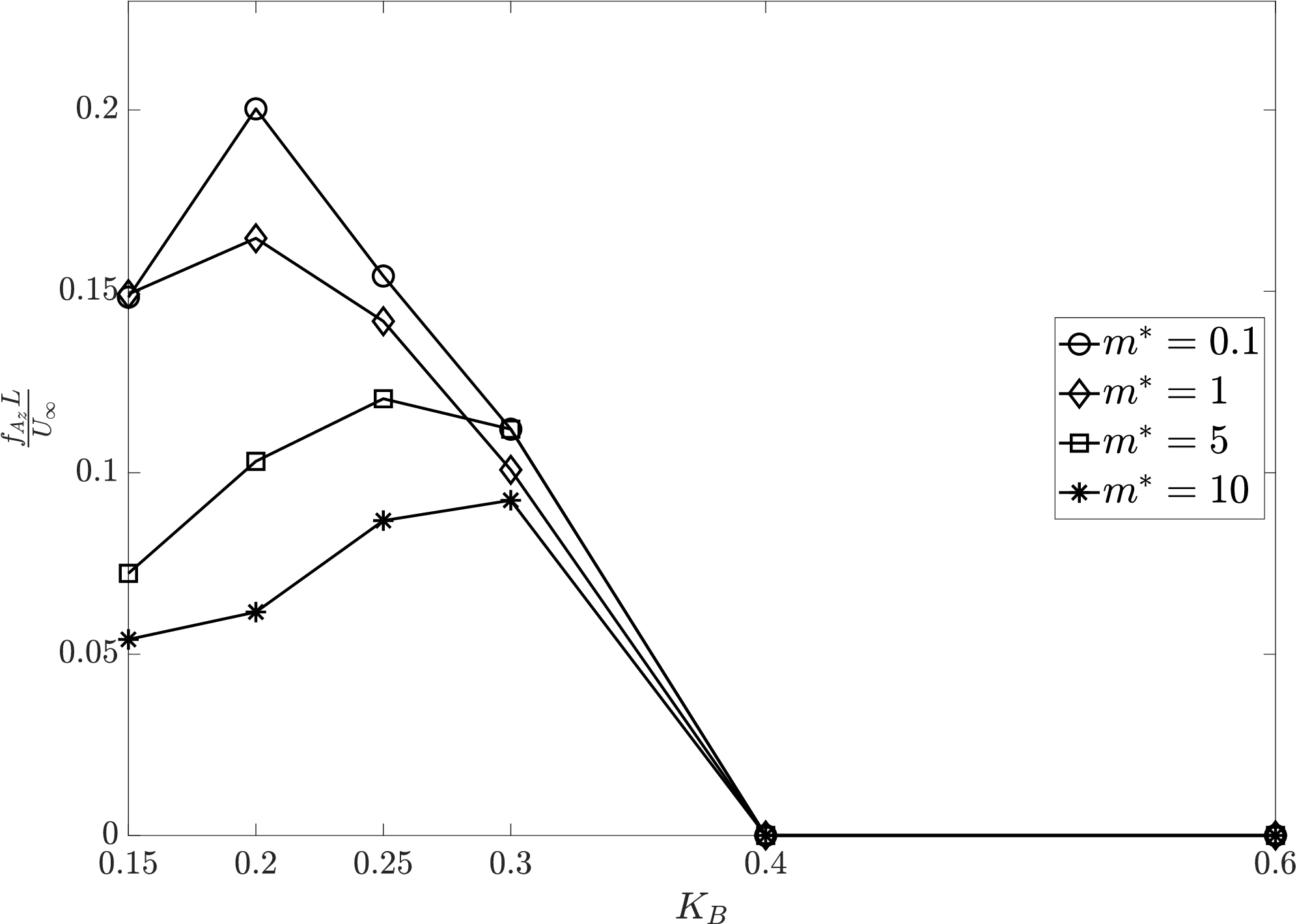}}
  \caption{Variation of flapping frequencies with non-dimensional bending stiffness $K_{B}\in[0.15, 1]$ and $m^{*}=[0.1,1,5,10]$ for an isolated inverted foil at $Re = 1000$}
\label{fig:10}
\end{figure}

\begin{figure}
 \centerline{\includegraphics[width=0.5\textwidth]{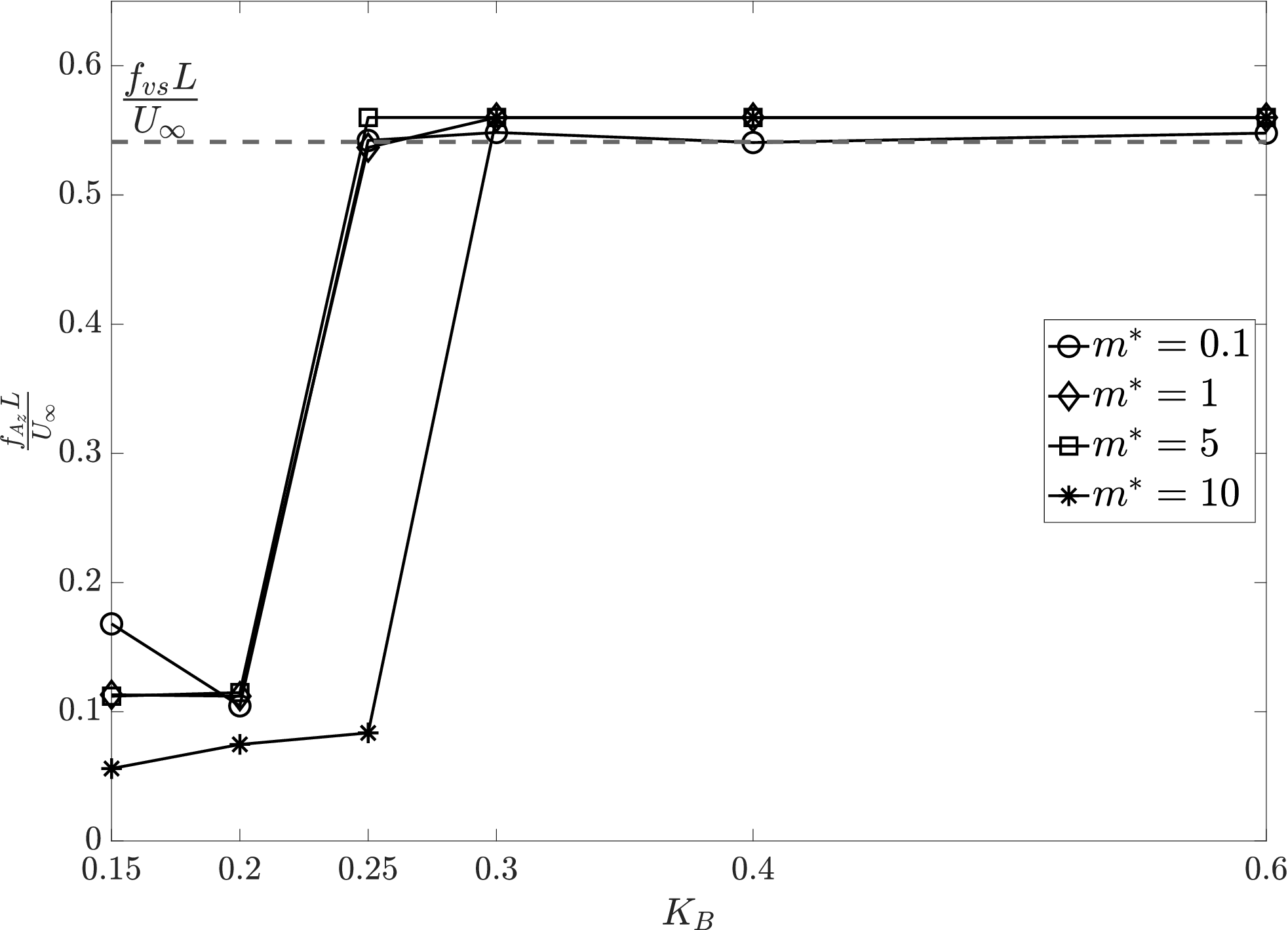}}
  \caption{Variation of flapping frequencies with non-dimensional bending stiffness $K_{B}\in[0.15, 1]$ and $m^{*}=[0.1,1,5,10]$ for the downstream inverted foil at $Re = 1000$}
\label{fig:11}
\end{figure}
We next investigate the influence of mass ratio on the inverted foil's response in the tandem setup. We analyze the response of the inverted foil over $K_{B}\in[0.15,  1]$, $m^{*} = [0.1, 1, 5, 10]$ and at a fixed $Re = 1000$. In our study, constant flow conditions are considered, therefore, the $m^{*}$ is varied by varying $\rho_{s}$. Figs. \ref{fig:8}-\ref{fig:9} show the variation maximum transverse tip amplitude for an inverted foil in the isolated and tandem setup as a function of $K_{B}$ and $m^{*}$. Figs. \ref{fig:10}-\ref{fig:11} show the variation of the transverse oscillation frequency as a function of $K_{B}$ and $m^{*}$ for an inverted foil in the isolated and tandem configurations. From Figs. \ref{fig:8}-\ref{fig:10}, we observe that the isolated foil becomes unstable below the same $K_{B}=0.35$ for all values of $m^*$ considered herein. Once unstable, the oscillation amplitude increases until it reaches peak amplitude as $m^*$ increases. Correspondingly, the oscillation frequency decreases as the foil becomes heavier. We attribute these variations in the flapping response to the increased inertia forces in the foil as it becomes heavier when $m^*$ is increased. Consequently, the isolated foil exhibits LAF response for a wider range of $K_{B}$ values at high $m^*$ compared to low $m^*$. These observations are in agreement with results reported by previous studies\cite{Gurugubelli2015Self-inducedFlow,Ryu2015FlappingFlow,Sader2016StabilityFlow}. 

In contrast to the isolated foil setup, under the influence of unsteady wake flow, the downstream foil's response shows a distinct behavior when the mass ratio is varied, as shown in Figs. \ref{fig:9}-\ref{fig:11}. We can broadly divide the response into two regimes based on the values of $m^{*}$ and $K_{B}$; (i) the dynamics are sensitive to both $m^{*}$ and $K_{B}$, and (ii) when the dynamics are influenced mainly by $K_{B}$. The first regime is observed when the downstream foil's dynamics are controlled by the cylinder vortex dynamics. During this regime, for $m^{*} = 0.1$, the foil undergoes small to moderate amplitude periodic flapping motion with oscillation frequency equal to the vortex shedding frequency of the upstream cylindrical bluff body as the $K_{B}$ is reduced. When $m^{*}$ is increased to $m^{*} = 1$, the MAF mode is not observed, and the foil oscillates in the SAF mode of all values of $K_{B}$. Upon further increase in $m^{*} = [5, 10]$, the flapping amplitudes decrease drastically and are almost negligible. It is worth noting that the flapping amplitude is synchronized with the lift force acting on the foil for all cases as presented in Fig. \ref{fig:9}.

Furthermore, in the first regime, for a given $K_{B}$, the flapping amplitude monotonically decreases with an increase in $m^{*}$. As a result, we observe that the rate at which the transverse tip displacement increases with $K_{B}$ decreases with an increase in $m^{*}$. We also observe that the $m^{*}$ influences the range in which the foil is synchronized with the cylinder vortex shedding. The range of fluid-structure parameters for which synchronization is observed is narrower for larger $m^{*}$ values as compared to the lower $m^{*}$ values. This indicates that in contrast to the isolated foil, the critical $K_{B,Cr}$ value below which the foil transitions to large amplitude flapping is influenced by $m^{*}$\cite{Gurugubelli2015Self-inducedFlow}. 
Compared to regime one, in the second regime, for all $m^{*}$ and $K_{B}$ values, the response is not synchronized with the cylinder vortex shedding and undergoes LAF with a single dominant frequency. As shown in Fig. \ref{fig:8}, when the mass ratio is increased, there is only a marginal increase in the transverse flapping amplitude, and a decrease in the flapping frequency with the mass ratio is observed, indicating that the response is mostly insensitive to changes in $m^{*}$. For a given mass ratio, the amplitude gradually increases with a decrease in $K_{B}$ until it reaches a peak value at $K_{B} = 0.15$. 

In general, these results suggest that while for an isolated foil, the dynamics are mostly insensitive to $m^*$ \cite{Gurugubelli2015Self-inducedFlow}, when subject to wake interference, the dynamics of the inverted foil are sensitive to both $K_{B}$ and $m^*$. For stiffer foils, at low mass ratios, i.e., when the foil is lighter, the higher loading provided by the unsteady wake of the bluff is sufficient in exciting the flapping response and maintaining the amplitudes. When $m^{*}$ is larger, and the foil becomes heavier, the inertia force in the foil increases and the unsteady loading from the wake is insufficient in balancing the inertia forces and maintaining the response amplitude. As a result even though the foil is synchronized with the vortex shedding, the amplitudes are minuscule compared to the lighter foils. For highly flexible foils $K_{B}< K_{B,Cr}$, the cylinder wake has limited influence on the downstream foil flapping dynamics. Moreover, the impact of the highly energetic wake further reduces when the foil becomes heavier, wherein the downstream foil de-synchronizes at a higher stiffness value than the lighter foils and undergoes flapping in the LAF-like mode observed in an isolated foil counterpart. This shows a relatively low sensitivity behavior towards large changes in mass ratio $m^{*}$ akin to an isolated foil \cite{kim2013flapping, Gurugubelli2015Self-inducedFlow}. In the LAF regime, similar to an isolated inverted foil when $m^{*}$ increases, the dampening effect on the foil's inertia forces by the wake is reduced, leading to a marginal increase in the flapping amplitude and a decrease in the frequency.


\begin{figure}
 \centerline{\includegraphics[width=0.4\textwidth]{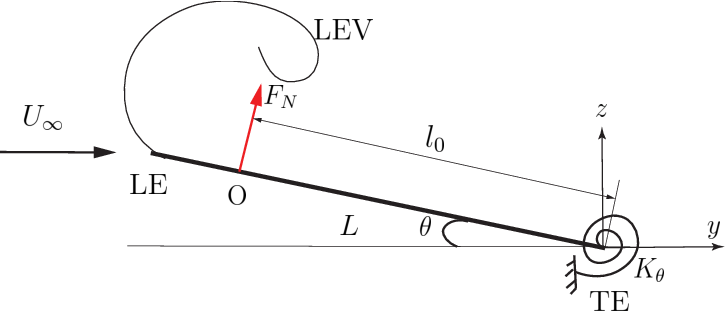}}
  \caption{Schematic of a flat plate cantilevered at the trailing edge with a torsional spring undergoing a single degree of freedom pitching rotation in uniform flow. }
\label{fig:schem_spring}
\end{figure}

\subsection{\label{sub:scale}Scaling relations for maximum transverse displacement}
As discussed in the previous Sec \ref{sub:Mass}, the impact of the unsteady wake on the response of the downstream inverted foil is strongly related to both the foil density and the elasticity for a given $Re$ and cylinder-foil geometry parameters. This indicates that some underlying mechanisms must exist between the foil's deformation and the non-dimensional bending rigidity and mass ratio. To quantify a relationship between them, we consider a simplified elastically-mounted model undergoing a single degree of freedom pitching motion or rotation in a uniform flow $U_{\infty}$, where $K_\theta$ is the torsional spring constant as shown in Fig. \ref{fig:schem_spring}. In this model the non-linear structural deformations of the inverted foil are simplified by assuming the flexible foil as a rigid plate of identical mass. The elastic nature of the foil is modeled by assuming an equivalent linear torsional spring. The fluid loads acting on the foil are represented by potential flow lift and the empirical drag based on the projected plate height. 
The equation of motion of the elastically-mounted rotating plate for the rotation angle $(\theta^{s})$ can be written as
\begin{equation}
    I\theta^{s}_{tt} + C_{\theta}\theta^{s}_{t} + K_{\theta}\theta^{s} = M^{f}_{x}, \label{eq:Mot}
\end{equation}
where $\theta^{s}_{tt}$ and $\theta^{s}_{t}$ represent $\partial^2 \theta^{s}/\partial t^2$ and $\partial \theta^{s}/\partial t$ respectively, $C_{\theta}$ is the torsional damping constant, $I$ is the mass moment of inertia for the rigid plate rotating about TE. $M^{f}_{x}$ denotes the steady moment acting on the plate due to the fluid loading acting normal to the plate at the aerodynamic center $\mathrm{O}$, where the pitching moment does not vary as a function of the twist angle. Here $l_{0}$ is the distance between the aerodynamic center and the trailing edge.
Consider a quasi-static loading on the rotating plate, the normal force via potential flow model and the force decomposition assumption is given as
\begin{equation}
    F_{N} = \frac{1}{2}\rho^{f}U^{2}_{\infty}L(C_{d}\mathrm{sin\theta^s|sin \theta^s|} + C_{l}\mathrm{cos\theta^s}),
    \label{eq:forcedecomp}
\end{equation}
where $C_{l}$ is the potential flow lift for the inclined flat plate in uniform flow at the angle of incidence $\theta^s$ and $C_{d}$ represents the drag coefficient of the flat plate in cross-flow. In equation \ref{eq:forcedecomp}, the terms $\mathrm{sin\theta^s}$ and $\mathrm{cos \theta^s}$ project the lift and drag force normal to the plate. The term $|\mathrm{sin \theta^s}|$ projects the inclined length of the plate in a direction normal to the incoming flow, while the absolute operator accounts for the sign of projected length. In the above analytical form, we assume, based on potential flow theory, that the flow remains attached to the plate even for a large rotation angle. The drag force and the inviscid inertia force are treated independently, based on Lighthill's force decomposition \cite{lighthill1986informal}. 
In order to separate the contribution due to structural forces and the fluid force, we non-dimensionalize time using flow variables $U_\infty$, $L$ and obtain the below relations:
\begin{equation}
\begin{array}{cc}
     &    t^{*} = tU_{\infty}/L;  \\
     &    \partial/\partial t = (\partial t^{*}/\partial t) (\partial/\partial t^{*}) = (U_{\infty}/L) \partial/\partial t^{*}
\end{array}
\end{equation}
by substituting in equation \ref{eq:Mot} we get:
\begin{equation}
    I U^{2}_{\infty}/L^2\theta^{s}_{t^{*}t^{*}} + C_{\theta}U_{\infty}/L\theta^{s}_{t^{*}} + K_{\theta}\theta^{s} (t^{*}) = M^{f}_{x}(t^{*}),\label{eq:Mot2}
\end{equation}
Considering a foil of unit width and dividing the whole equation by $\rho^{f}U^{2}_{\infty}L^2$ we obtain:
\begin{equation}
    M^{*}\theta^{s}_{t^{*}t^{*}} + C^{*}_{\theta}\theta^{s}_{t^{*}}  + K^{*}_{\theta}\theta^{s}(t^{*}) =M^{f^{*}}_{x}(t^{*}) =  C_n(t^{*})l^{*}, 
    \label{eq:main}
\end{equation}
where
\begin{equation}
  M^{*} = \frac{I}{\frac{1}{2}\rho^{f}L^4},\;  C^{*}_{\theta} =\frac{C_{\theta}}{\frac{1}{2}\rho^{f}U_{\infty}L^3}, \;K^{*}_{\theta} = \frac{K_{\theta}}{\frac{1}{2}\rho^{f}U^{2}_{\infty}L^2},
\end{equation}
here $l^{*} = \frac{l_{0}}{L},$ is the non-dimensional distance of the aerodynamic center from the trailing edge of the foil, and $C_n$ is the coefficient of the normal force acting on the foil. Now, apart from these non-dimensional parameters, the dynamics of the system are also dependent on the Reynolds number $Re$; however, for our study, we consider a constant $Re = 1000$.

Consider that the foil undergoes a simple sinusoidal oscillation $\theta^{s} (t^{*})= \theta_{0}\mathrm{sin}(\frac{\omega^*}{2\pi}t^{*})$, and the fluid momentum with a phase ${\phi}$, $M^{f^{*}}_{x}(t^{*}) = M^{f}_{x_0}\mathrm{sin}(\frac{\omega^*}{2\pi}t^{*} +\phi)$, equation \ref{eq:main} can be rewritten as:
\begin{equation}
\begin{array}{ll}
     &-M^{*}\left(\frac{\omega^*}{2\pi}\right)^2\theta_{0}\mathrm{sin}(\frac{\omega^*}{2\pi}t^{*}) +C^{*}_{\theta}\left(\frac{\omega^*}{{2\pi}}\right)\theta_0 \mathrm{cos}(\frac{\omega^*}{2\pi}t^{*})\\
     &+K^{*}_{\theta}\theta_{0}\mathrm{sin}(\frac{\omega^*}{2\pi}t^{*}) =M^{f^{*}}_{x_0}\mathrm{sin}(\frac{\omega^*}{2\pi}t^{*})\mathrm{cos}(\phi) \\
     &+M^{f^{*}}_{x_0}\mathrm{cos}(\frac{\omega^*}{2\pi}t^{*})\mathrm{sin}(\phi),
\end{array}
\end{equation}
where $\frac{\omega^*}{2\pi} = f_{\theta}^{*}$, $f_{\theta}^{*} = f_L/U_{\infty}$ is the non-dimensional oscillation frequency of the plate,  and $\theta_{0}$ is angular displacement. Next, the coefficients in front of $\mathrm{sin}(\frac{\omega^*}{2\pi}t^{*})$ and $\mathrm{cos}(\frac{\omega^*}{2\pi}t^{*})$ can be equated as follows:
    \begin{align}
         & (-M^{*}f_{\theta}^{*^2} + K^{*}_{\theta}) \theta_{0} = M^{f^{*}}_{x_0}\mathrm{cos}(\phi),  \\
         & C^{*}_{\theta}f_{\theta}^{*}\theta_{0} = M^{f^{*}}_{x_0}\mathrm{sin}(\phi).   
    \end{align}
This results in a new non-dimensional parameter called effective stiffness $K_{eff} =-M^{*}f_{\theta}^{*^2} + K^{*}_{\theta}$. This effective stiffness parameter is defined in a manner similar to the effective stiffness parameter introduced by \cite{shiels2001flow} and \cite{gharib1997exploration} for an elastically mounted cylinder subject to a uniform flow.  
Now consider if the motion of the plate is in-phase with the fluid loading i.e., $\phi = 0$, the effective stiffness relates the steady moment acting on the plate $M^f_{x}$ to $\theta_{0}$ ($M^{f}_{x} = K_{eff}\theta_{0}$) akin to a spring stiffness $k$ for the external force $F$ to spring displacement $z$ $(F = -kz)$. This relation also shows that for an undamped system, the response of the flat plate under quasi-static loading can be described based on the effective stiffness parameter. Moreover, the periodic motion of the plate is sustained by the balance between the fluid loading and the structure's inertia and elastic forces.
Similar to the response of a flat plate, if a flexible foil is inverted,i.e., fixed at the leading edge, we observe that it also undergoes self-sustaining periodic flapping motion and experiences periodic fluid loading when subjected to a uniform flow. 

Given a fixed Reynolds number and negligible damping, the response of the flexible foil can be characterized by the non-dimensional bending rigidity $K_{B}$ and the structure-to-fluid mass ratio $m^{*}$. Therefore, we propose an effective stiffness parameter $K^{*}_{eff}$ based on the effective stiffness for a rigid plate undergoing a rotational-pitching motion under fluid flow as follows:
\begin{equation}
    K^{*}_{eff} = K_{B} - f^{*^2}m^{*},
    \label{eq:effK}
\end{equation}
where $f^{*} = f_{A_{z}}/U_{\infty}$ is the non-dimensional frequency of the transverse tip displacement of the flexible inverted foil. Here, we consider that the fluid loading and foil's motion is in-phase. Through this parameter, we can combine the effect of the foil's elasticity and inertia forces on the dynamics. Moreover, for an undamped system, this parameter expresses the fact that for pure sinusoidal motion, the body's inertia force essentially opposes the resistive elastic force. 
\begin{figure}
 \centerline{\includegraphics[width=0.47\textwidth]{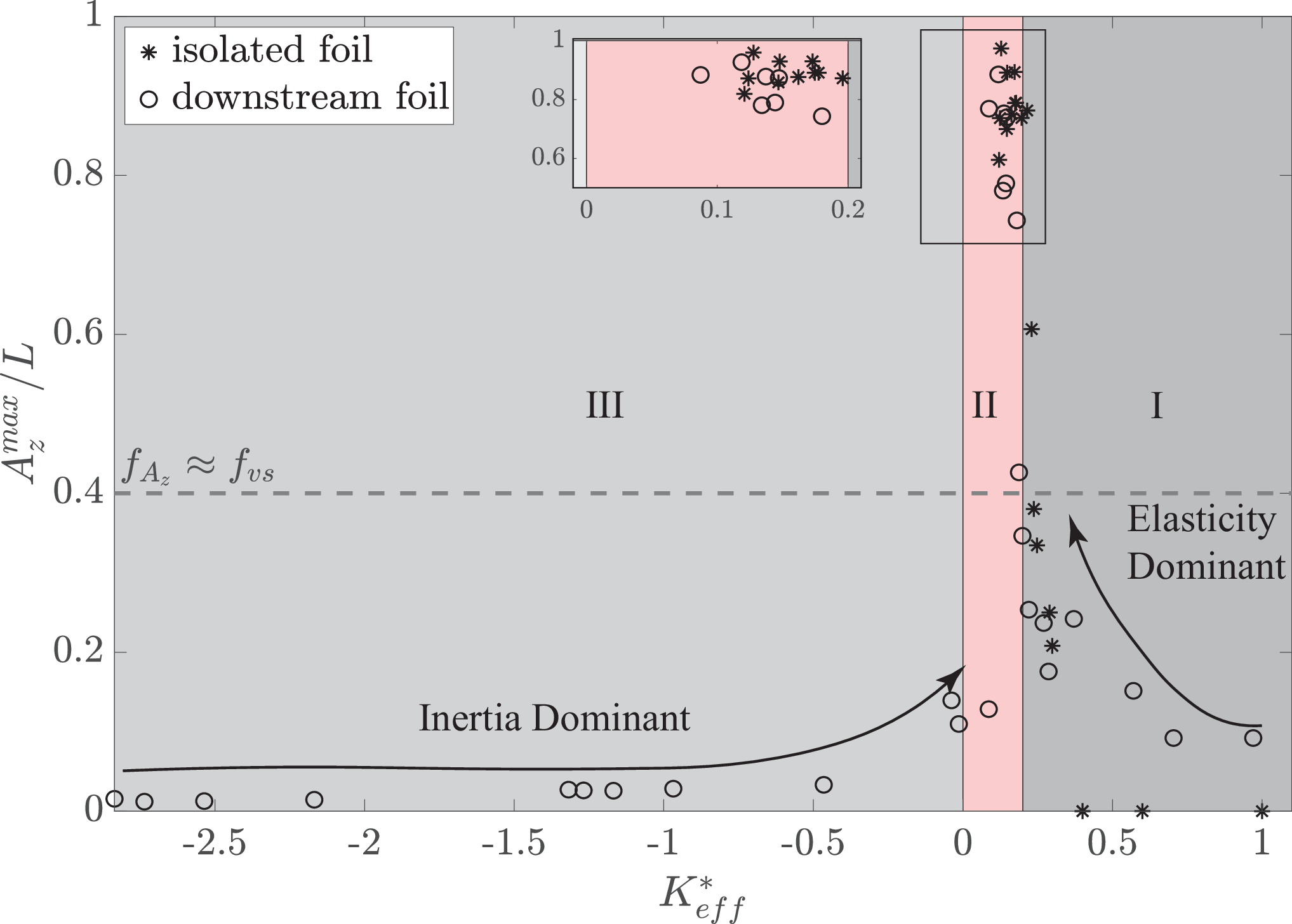}}
  \caption{Variation of the maximum transverse tip displacement of the isolated $(*)$ and the downstream foil $(\circ)$ foil with $K^*_{eff}$. Here the dotted line (- -) represents the $A^{max}_{z}/L$ below which synchronization is observed.}
\label{fig:scaling}
\end{figure}

Figure \ref{fig:scaling} represents the variation of maximum transverse tip displacement of the inverted foil in the isolated and tandem setups as a function of this new parameter $K^{*}_{eff}$. Here positive $K^{*}_{eff}$ values represent the inverted foils with small $m^{*}$ values or light foils therefore, $K_{B}> f^{*^2}m^{*}$ while negative $K^{*}_{eff}$ represents the cases with large $m^{*}$ values, thus $K_{B} < f^{*^2}m^{*}$.  If we consider an isolated inverted foil, as $K^{*}_{eff}$ is decreased, we observe that up to a critical value of $K^{*}_{eff} = 0.4$, the isolated foil is stable. Below this critical value, the foil rapidly experiences starts flapping with large amplitudes. As shown in the close-up subfigure in Fig. \ref {fig:scaling}, for a range of $0.1<K^{*}_{eff}<0.2$ values, minor variations in amplitudes are recorded for the isolated as $K^{*}_{eff}$ is reduced.  For this range of $K^{*}_{eff}$ values, the inverted foil undergoes large amplitude oscillations with low frequencies. As a result of the low flapping frequency, the large variations in $m^{*}$ result in only a minor variation in $K_{B}$ and $K^{*}_{eff}\approx K_{B}$. As a result, large variations in $m^{*}$ do not affect the range $K_{B}$ in which flapping is observed earlier in Fig. \ref{fig:9}. These results provide further evidence that the dynamics for an isolated inverted foil are dependent mainly on the structure's elasticity while remain mostly insensitive to the large changes in $m^*$ \cite{kim2013flapping,Gurugubelli2015Self-inducedFlow,Gurugubelli2019LargePeriodicity}.

In contrast to the isolated foil, for the downstream foil, two branches are identified based on the $K^{*}_{eff}$ through which the foil transitions to large amplitude flapping. The upper branch is observed for $K^{*}_{eff} > 0.2$ while the lower branch is observed for $K^{*}_{eff}< 0$.  It is important to note that both branches represent the response of the downstream foil when it is synchronized with the cylinder vortex shedding; thus, $f^{*}$ is nearly constant. As shown in Fig. \ref{fig:scaling},  the downstream foil first undergoes SAF followed by MAF and gradually transitions to the LAF regime for the upper branch. However, for the lower branch, the downstream foil only undergoes SAF with a gradual increase in oscillation amplitude as $K^{*}_{eff}$ is increased until it. Compared to the upper branch, the lower branch transitions to the large amplitude regime at a faster rate as compared to the upper branch. 

As observed earlier in the tandem setup, the flapping response of the downstream foil is highly correlated to both $m^{*}$ and $K_{B}$ during synchronization. The transition to LAF is more gradual for light foils, compared to heavy foil. Both these trends are accurately represented by $K^*_{eff}$ through the upper and lower branches. For $0< K^{*}_{eff}< 0.2$, the foil undergoes low-frequency large-amplitude flapping response that is mostly independent to $m^{*}$ similar to an isolated inverted foil as observed in Fig. \ref{fig:scaling}.

Based on these observations, we characterize the response of an inverted foil in the tandem setup into three zones as a function of decreasing $K^{*}_{eff}$; Zone $\mathrm{I}$, Zone $\mathrm{II}$ and Zone $\mathrm{III}$. Zone ${\mathrm{I}}$ is observed for $K^{*}_{eff}>0.2$; it characterizes the region where the inverted foil undergoes wake-induced oscillations with small to moderate amplitudes wherein the oscillations are driven by the feedback between fluid loading due to wake vortices and dominantly foil's internal elastic forces. Zone $\mathrm{III}$ is observed for $K^{*}_{eff}<0$; it characterizes the region where the inverted foil undergoes wake-induced oscillations with small to moderate amplitudes wherein the oscillations are driven by the feedback between fluid loading due to wake vortices and dominantly foil's inertia forces.  Lastly, Zone $\mathrm{II}$ is observed for $0<K^{*}_{eff}<0.2$, in which no synchronization is observed, and the downstream foil undergoes LAF motion similar to an isolated foil.
Overall we observe that the $K^{*}_{eff}$ parameter, indeed, collapses the two parameters $m^{*}$ and ${K_{B}}$ into one parameter, which provides a unified scaling of the system and captures the varied dynamics of inverted foils the tandem setup. Given the motion is dominated by a single frequency, as demonstrated earlier Fig. \ref{fig:10}. 

\section{\label{sec:conclusions} Conclusions}
In this work, we have investigated the response of an inverted foil when placed in tandem with a stationary circular cylinder using high-fidelity numerical simulations. The dynamics are explored systematically as a function of the non-dimensional bending stiffness and mass ratio at a fixed Reynolds number and the gap ratio. Our analysis demonstrates that due to interaction with the cylinder's unsteady wake, the downstream becomes unstable for a higher stiffness value than the isolated case due to the higher transverse fluid loading resulting from the unsteady wake flow. Consequently, it undergoes sustained oscillations for a wider range of the non-dimensional bending stiffness. As a function of increasing flexibility, two additional flapping modes are observed for the downstream foil before transitioning to large amplitude flapping mode: (i) small amplitude flapping and (ii) moderate amplitude flapping modes.

A critical non-dimensional bending stiffness value of $K_{B, Cr} = 0.25$ exists above which synchronization between the downstream foil and cylinder vortex shedding is observed, which drives the foil's flapping response. During synchronization, the inverted foil oscillates with small to moderate amplitudes as stiffness is reduced.  Also, a C(2S) shedding pattern wake is observed, which quickly develops to a 2S pattern in the near wake of the foil. Below this threshold stiffness, the cylinder loses control over the foil, and the downstream foil undergoes sustained flapping like an isolated foil. 

Upon investigating the downstream inverted foil dynamics at different mass ratios, our results indicated that, compared to uniform flow, when subject to the unsteady wake of a bluff body, the wake interference effects on the foil, along with the occurrence of various flapping modes, are highly influenced by the mass ratio. We can characterize the foil's response-based function of $m^{*}$ into two regimes. In the first region, the foil's motion is highly sensitive to $m^{*}$ and negatively correlated to $m^{*}$ and $K_{B}$. This region is observed when the foil's motion is synchronized with the cylinder.  Apart from influencing the foil's transverse motion, $m^{*}$ also influences the stability boundary of this zone where in the threshold $K_{B}$ value at which the foil desynchronizes and transitions to the LAF regime increases with $m^{*}$. In the second region, we observe that the downstream foil is almost insensitive to $m^{*}$ and undergoes low-frequency LAF similar to an isolated foil. 

Based on these observations, we propose a new effective stiffness parameter that captures the combined effect of $K_{B}$ and $m^{*}$ on the dynamics of inverted foils in these setups. This parameter is derived based on the governing equation of a 2D rigid flat plate undergoing a single degree of freedom pitching motion or rotation in uniform flow and an empirical flow model. This new parameter revealed two branches that describe the downstream foil's response when it is under the cylinder's influence: (i) the lower branch and (ii) the upper branch. The lower branch describes the variation in foil's amplitude when its response is majorly driven by inertia forces i.e., for heavy foils (large $m^*$), while the upper branch describes it for the cases when the oscillations are sustained by the foil's elastic forces. These results can facilitate the development of novel optimization and control concepts that maximize the energy-harvesting potential of inverted foils as well as ensure a consistent energy output. Eventually, these control strategies can be leveraged for the design of efficient and robust inverted foil-based energy-harvesting array systems.

\begin{acknowledgments}
The authors would like to acknowledge the Natural Sciences and Engineering Research Council of Canada (NSERC) for funding the project. The research was enabled in part through computational resources and services provided by Compute Canada, and the Advanced Research Computing facility at the University of British Columbia.
\end{acknowledgments}

\appendix
\section{\label{sect:Appendix}Validation} 
\begin{figure}[h]
 \centerline{\includegraphics[width=0.5\textwidth]{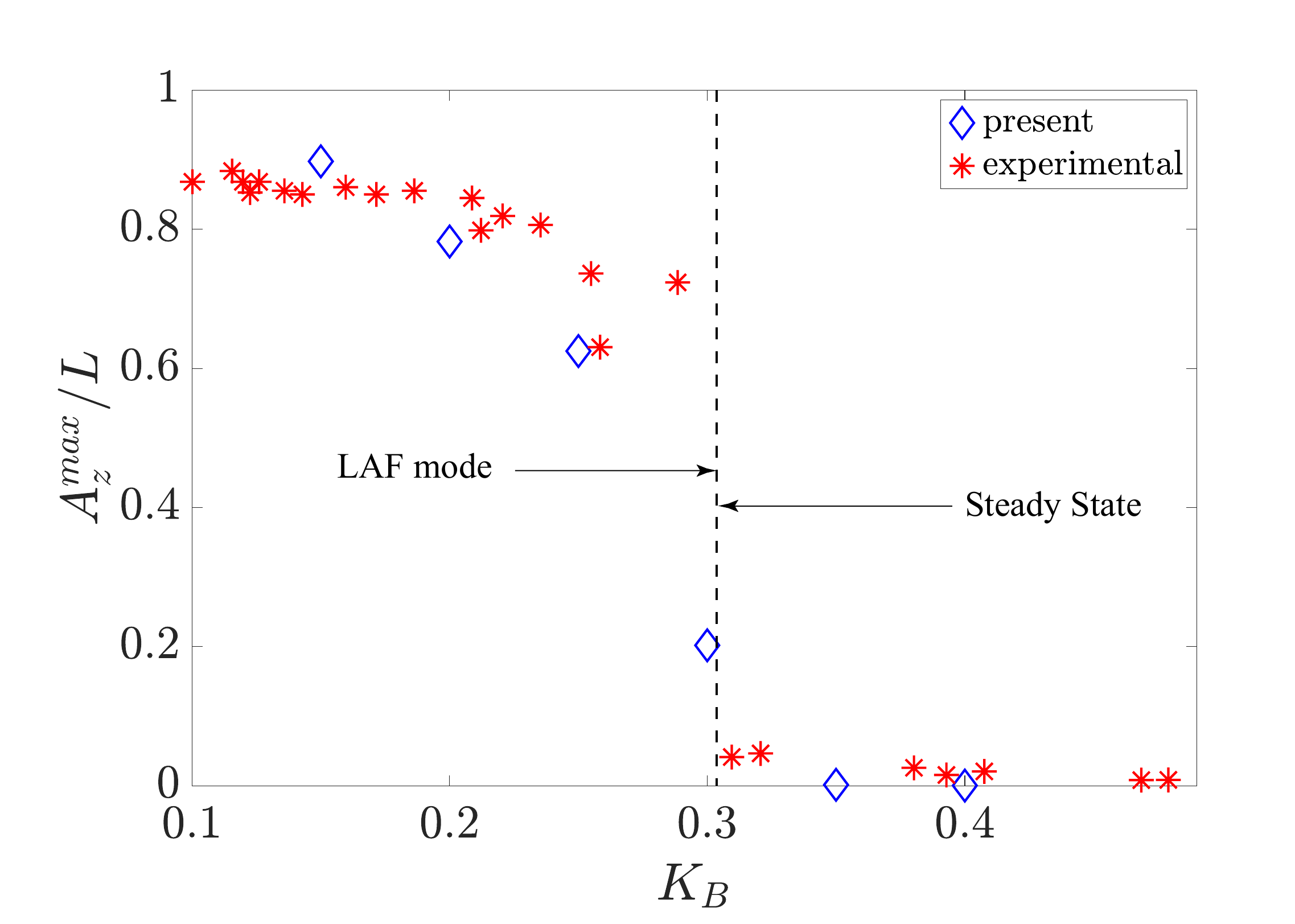}}
  \caption{Comparison of the maximum transverse tip displacement of an isolated inverted foil between the present data and experimental results by \cite{kim2013flapping} at $m^* = 1$}
\label{fig:KA}
\end{figure}

\begin{figure}[h]
 \centerline{\includegraphics[width=0.47\textwidth]{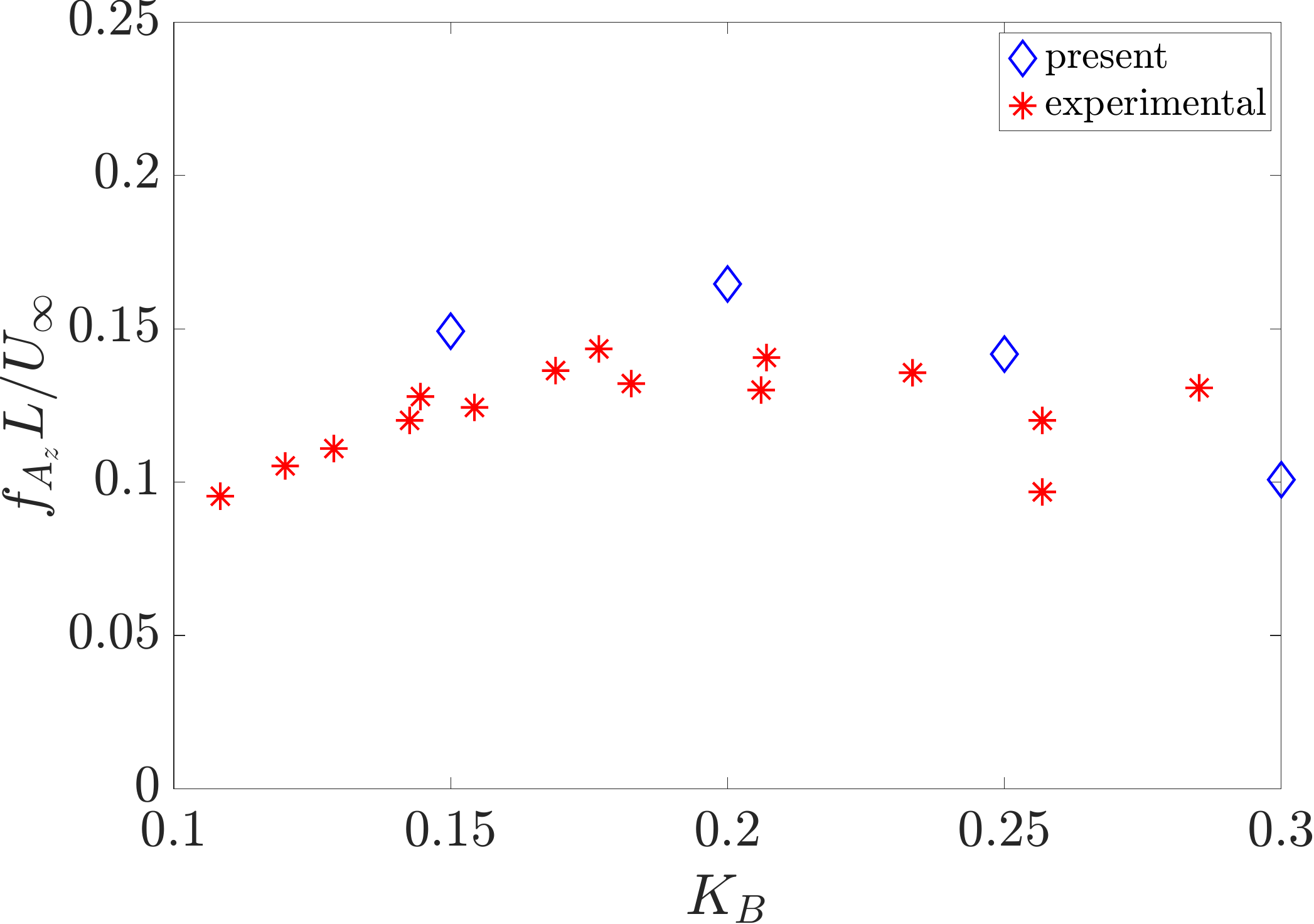}}
  \caption{Comparison of the dominant transverse tip displacement frequency of an isolated inverted foil between the present data and the experimental results of \cite{kim2013flapping} at $m^* = 1$}
\label{fig:KAf}
\end{figure}

The 2D computational setup established in Sec \ref{sect:Compdomain} is considered for our comparative study and validation purposes. We first consider the case of an inverted foil in the isolated setup, i.e., without the upstream cylinder for the non-dimensional parameters $K_{B} \in [0.15, 0.5]$ and $m^{*} =1$. These parameters are consistent with the experimental conditions of \cite{kim2013flapping} for $m^* = O(1)$. To contrast the present study with experimental results, figures \ref{fig:KA} and \ref{fig:KAf} show the comparison of the variation in the maximum transverse tip displacement and the dominant flapping frequency with non-dimensional bending rigidity $K_{B}$.  These plots confirm that our numerical simulations can correctly predict the onset of flapping instability and effectively capture the variations in $A^{max}_{z}/L$ and $f_{A_{z}}L/U_\infty$ in the large-amplitude flapping regime.  Some discrepancies are observed in $A^{max}_{z}/L$ and $f_{A_{z}}L/U_\infty$  obtained in the present study and the experimental results. We attribute these differences to the three-dimensional effects, which have not been considered for this numerical study, while the flexible foils used by \cite{kim2013flapping} were of aspect ratios in the range of $1-1.3$. We also attribute these discrepancies in both setups to the large difference in Reynolds numbers of the experimental studies and the present numerical investigation. 



\bibliography{ref}

\providecommand{\noopsort}[1]{}\providecommand{\singleletter}[1]{#1}%
\begin{thebibliography}{53}%
\makeatletter
\providecommand \@ifxundefined [1]{%
 \@ifx{#1\undefined}
}%
\providecommand \@ifnum [1]{%
 \ifnum #1\expandafter \@firstoftwo
 \else \expandafter \@secondoftwo
 \fi
}%
\providecommand \@ifx [1]{%
 \ifx #1\expandafter \@firstoftwo
 \else \expandafter \@secondoftwo
 \fi
}%
\providecommand \natexlab [1]{#1}%
\providecommand \enquote  [1]{``#1''}%
\providecommand \bibnamefont  [1]{#1}%
\providecommand \bibfnamefont [1]{#1}%
\providecommand \citenamefont [1]{#1}%
\providecommand \href@noop [0]{\@secondoftwo}%
\providecommand \href [0]{\begingroup \@sanitize@url \@href}%
\providecommand \@href[1]{\@@startlink{#1}\@@href}%
\providecommand \@@href[1]{\endgroup#1\@@endlink}%
\providecommand \@sanitize@url [0]{\catcode `\\12\catcode `\$12\catcode `\&12\catcode `\#12\catcode `\^12\catcode `\_12\catcode `\%12\relax}%
\providecommand \@@startlink[1]{}%
\providecommand \@@endlink[0]{}%
\providecommand \url  [0]{\begingroup\@sanitize@url \@url }%
\providecommand \@url [1]{\endgroup\@href {#1}{\urlprefix }}%
\providecommand \urlprefix  [0]{URL }%
\providecommand \Eprint [0]{\href }%
\providecommand \doibase [0]{https://doi.org/}%
\providecommand \selectlanguage [0]{\@gobble}%
\providecommand \bibinfo  [0]{\@secondoftwo}%
\providecommand \bibfield  [0]{\@secondoftwo}%
\providecommand \translation [1]{[#1]}%
\providecommand \BibitemOpen [0]{}%
\providecommand \bibitemStop [0]{}%
\providecommand \bibitemNoStop [0]{.\EOS\space}%
\providecommand \EOS [0]{\spacefactor3000\relax}%
\providecommand \BibitemShut  [1]{\csname bibitem#1\endcsname}%
\let\auto@bib@innerbib\@empty
\bibitem [{\citenamefont {Watanabe}\ \emph {et~al.}(2002)\citenamefont {Watanabe}, \citenamefont {Isogai}, \citenamefont {Suzuki},\ and\ \citenamefont {Sugihara}}]{Watanabe2002AFlutter}%
  \BibitemOpen
  \bibfield  {author} {\bibinfo {author} {\bibfnamefont {Y.}~\bibnamefont {Watanabe}}, \bibinfo {author} {\bibfnamefont {K.}~\bibnamefont {Isogai}}, \bibinfo {author} {\bibfnamefont {S.}~\bibnamefont {Suzuki}},\ and\ \bibinfo {author} {\bibfnamefont {M.}~\bibnamefont {Sugihara}},\ }\bibfield  {title} {\bibinfo {title} {{A theoretical study of paper flutter}},\ }\href {https://doi.org/10.1006/jfls.2001.0436} {\bibfield  {journal} {\bibinfo  {journal} {Journal of Fluids and Structures}\ }\textbf {\bibinfo {volume} {16}},\ \bibinfo {pages} {543} (\bibinfo {year} {2002})}\BibitemShut {NoStop}%
\bibitem [{\citenamefont {Guo}\ \emph {et~al.}(2000)\citenamefont {Guo}, \citenamefont {Paidoussis},\ and\ \citenamefont {Asme}}]{Guo2000StabilityFlow}%
  \BibitemOpen
  \bibfield  {author} {\bibinfo {author} {\bibfnamefont {C.~Q.}\ \bibnamefont {Guo}}, \bibinfo {author} {\bibfnamefont {M.~P.}\ \bibnamefont {Paidoussis}},\ and\ \bibinfo {author} {\bibfnamefont {F.}~\bibnamefont {Asme}},\ }\href {http://asmedigitalcollection.asme.org/appliedmechanics/article-pdf/67/1/171/5466487/171_1.pdf} {\emph {\bibinfo {title} {{Stability of Rectangular Plates With Free Side-Edges in Two-Dimensional Inviscid Channel Flow}}}},\ \bibinfo {type} {Tech. Rep.}\ (\bibinfo {year} {2000})\BibitemShut {NoStop}%
\bibitem [{\citenamefont {Allen}\ and\ \citenamefont {Smits}(2001)}]{Allen2001EnergyEel}%
  \BibitemOpen
  \bibfield  {author} {\bibinfo {author} {\bibfnamefont {J.~J.}\ \bibnamefont {Allen}}\ and\ \bibinfo {author} {\bibfnamefont {A.~J.}\ \bibnamefont {Smits}},\ }\bibfield  {title} {\bibinfo {title} {{Energy Harvesting Eel}},\ }\href {https://doi.org/10.1006/j{\#}s.2000.0355} {\bibfield  {journal} {\bibinfo  {journal} {Journal of Fluids and Structures}\ }\textbf {\bibinfo {volume} {15}},\ \bibinfo {pages} {629} (\bibinfo {year} {2001})}\BibitemShut {NoStop}%
\bibitem [{\citenamefont {Kim}\ \emph {et~al.}(2013)\citenamefont {Kim}, \citenamefont {Coss{\'e}}, \citenamefont {Cerdeira},\ and\ \citenamefont {Gharib}}]{kim2013flapping}%
  \BibitemOpen
  \bibfield  {author} {\bibinfo {author} {\bibfnamefont {D.}~\bibnamefont {Kim}}, \bibinfo {author} {\bibfnamefont {J.}~\bibnamefont {Coss{\'e}}}, \bibinfo {author} {\bibfnamefont {C.~H.}\ \bibnamefont {Cerdeira}},\ and\ \bibinfo {author} {\bibfnamefont {M.}~\bibnamefont {Gharib}},\ }\bibfield  {title} {\bibinfo {title} {Flapping dynamics of an inverted flag},\ }\href@noop {} {\bibfield  {journal} {\bibinfo  {journal} {Journal of Fluid Mechanics}\ }\textbf {\bibinfo {volume} {736}},\ \bibinfo {pages} {R1} (\bibinfo {year} {2013})}\BibitemShut {NoStop}%
\bibitem [{\citenamefont {Orrego}\ \emph {et~al.}(2017)\citenamefont {Orrego}, \citenamefont {Shoele}, \citenamefont {Ruas}, \citenamefont {Doran}, \citenamefont {Caggiano}, \citenamefont {Mittal},\ and\ \citenamefont {Kang}}]{Orrego2017HarvestingFlag}%
  \BibitemOpen
  \bibfield  {author} {\bibinfo {author} {\bibfnamefont {S.}~\bibnamefont {Orrego}}, \bibinfo {author} {\bibfnamefont {K.}~\bibnamefont {Shoele}}, \bibinfo {author} {\bibfnamefont {A.}~\bibnamefont {Ruas}}, \bibinfo {author} {\bibfnamefont {K.}~\bibnamefont {Doran}}, \bibinfo {author} {\bibfnamefont {B.}~\bibnamefont {Caggiano}}, \bibinfo {author} {\bibfnamefont {R.}~\bibnamefont {Mittal}},\ and\ \bibinfo {author} {\bibfnamefont {S.~H.}\ \bibnamefont {Kang}},\ }\bibfield  {title} {\bibinfo {title} {{Harvesting ambient wind energy with an inverted piezoelectric flag}},\ }\href {https://doi.org/10.1016/j.apenergy.2017.03.016} {\bibfield  {journal} {\bibinfo  {journal} {Applied Energy}\ }\textbf {\bibinfo {volume} {194}},\ \bibinfo {pages} {212} (\bibinfo {year} {2017})}\BibitemShut {NoStop}%
\bibitem [{\citenamefont {Ribeiro}\ \emph {et~al.}(2021)\citenamefont {Ribeiro}, \citenamefont {Su}, \citenamefont {Guillaumin}, \citenamefont {Breuer},\ and\ \citenamefont {Franck}}]{ribeiro2021wake}%
  \BibitemOpen
  \bibfield  {author} {\bibinfo {author} {\bibfnamefont {B.~L.~R.}\ \bibnamefont {Ribeiro}}, \bibinfo {author} {\bibfnamefont {Y.}~\bibnamefont {Su}}, \bibinfo {author} {\bibfnamefont {Q.}~\bibnamefont {Guillaumin}}, \bibinfo {author} {\bibfnamefont {K.~S.}\ \bibnamefont {Breuer}},\ and\ \bibinfo {author} {\bibfnamefont {J.~A.}\ \bibnamefont {Franck}},\ }\bibfield  {title} {\bibinfo {title} {Wake-foil interactions and energy harvesting efficiency in tandem oscillating foils},\ }\href@noop {} {\bibfield  {journal} {\bibinfo  {journal} {Physical Review Fluids}\ }\textbf {\bibinfo {volume} {6}},\ \bibinfo {pages} {074703} (\bibinfo {year} {2021})}\BibitemShut {NoStop}%
\bibitem [{\citenamefont {Math{\'{u}}na}\ \emph {et~al.}(2008)\citenamefont {Math{\'{u}}na}, \citenamefont {O'Donnell}, \citenamefont {Martinez-Catala}, \citenamefont {Rohan},\ and\ \citenamefont {O'Flynn}}]{Mathuna2008EnergyNetworks}%
  \BibitemOpen
  \bibfield  {author} {\bibinfo {author} {\bibfnamefont {C.~Ã.}\ \bibnamefont {Math{\'{u}}na}}, \bibinfo {author} {\bibfnamefont {T.}~\bibnamefont {O'Donnell}}, \bibinfo {author} {\bibfnamefont {R.~V.}\ \bibnamefont {Martinez-Catala}}, \bibinfo {author} {\bibfnamefont {J.}~\bibnamefont {Rohan}},\ and\ \bibinfo {author} {\bibfnamefont {B.}~\bibnamefont {O'Flynn}},\ }\bibfield  {title} {\bibinfo {title} {{Energy scavenging for long-term deployable wireless sensor networks}},\ }\href {https://doi.org/10.1016/j.talanta.2007.12.021} {\bibfield  {journal} {\bibinfo  {journal} {Talanta}\ }\textbf {\bibinfo {volume} {75}},\ \bibinfo {pages} {613} (\bibinfo {year} {2008})}\BibitemShut {NoStop}%
\bibitem [{\citenamefont {Erturk}\ \emph {et~al.}(2010)\citenamefont {Erturk}, \citenamefont {Vieira}, \citenamefont {De~Marqui},\ and\ \citenamefont {Inman}}]{erturk2010energy}%
  \BibitemOpen
  \bibfield  {author} {\bibinfo {author} {\bibfnamefont {A.}~\bibnamefont {Erturk}}, \bibinfo {author} {\bibfnamefont {W.~G.~R.}\ \bibnamefont {Vieira}}, \bibinfo {author} {\bibfnamefont {C.}~\bibnamefont {De~Marqui}},\ and\ \bibinfo {author} {\bibfnamefont {D.~J.}\ \bibnamefont {Inman}},\ }\bibfield  {title} {\bibinfo {title} {On the energy harvesting potential of piezoaeroelastic systems},\ }\href@noop {} {\bibfield  {journal} {\bibinfo  {journal} {Applied physics letters}\ }\textbf {\bibinfo {volume} {96}} (\bibinfo {year} {2010})}\BibitemShut {NoStop}%
\bibitem [{\citenamefont {Erturk}\ and\ \citenamefont {Inman}(2011)}]{Erturk2011PiezoelectricHarvesting}%
  \BibitemOpen
  \bibfield  {author} {\bibinfo {author} {\bibfnamefont {A.}~\bibnamefont {Erturk}}\ and\ \bibinfo {author} {\bibfnamefont {D.~J.}\ \bibnamefont {Inman}},\ }\href@noop {} {\emph {\bibinfo {title} {{Piezoelectric energy harvesting}}}}\ (\bibinfo  {publisher} {John Wiley {\textbackslash}{\&} Sons},\ \bibinfo {year} {2011})\BibitemShut {NoStop}%
\bibitem [{\citenamefont {Taneda}(1968)}]{taneda1968waving}%
  \BibitemOpen
  \bibfield  {author} {\bibinfo {author} {\bibfnamefont {S.}~\bibnamefont {Taneda}},\ }\bibfield  {title} {\bibinfo {title} {Waving motions of flags},\ }\href@noop {} {\bibfield  {journal} {\bibinfo  {journal} {Journal of the Physical Society of Japan}\ }\textbf {\bibinfo {volume} {24}},\ \bibinfo {pages} {392} (\bibinfo {year} {1968})}\BibitemShut {NoStop}%
\bibitem [{\citenamefont {Argentina}\ and\ \citenamefont {Mahadevan}(2005)}]{Argentina2005Fluid-flow-inducedFlag}%
  \BibitemOpen
  \bibfield  {author} {\bibinfo {author} {\bibfnamefont {M.}~\bibnamefont {Argentina}}\ and\ \bibinfo {author} {\bibfnamefont {L.}~\bibnamefont {Mahadevan}},\ }\bibfield  {title} {\bibinfo {title} {{Fluid-flow-induced flutter of a flag}},\ }\href {https://doi.org/10.1073/pnas.0408383102} {\bibfield  {journal} {\bibinfo  {journal} {Proceedings of the National Academy of Sciences of the United States of America}\ }\textbf {\bibinfo {volume} {102}},\ \bibinfo {pages} {1829} (\bibinfo {year} {2005})}\BibitemShut {NoStop}%
\bibitem [{\citenamefont {Akcabay}\ and\ \citenamefont {Young}(2012)}]{Akcabay2012HydroelasticFlow}%
  \BibitemOpen
  \bibfield  {author} {\bibinfo {author} {\bibfnamefont {D.~T.}\ \bibnamefont {Akcabay}}\ and\ \bibinfo {author} {\bibfnamefont {Y.~L.}\ \bibnamefont {Young}},\ }\bibfield  {title} {\bibinfo {title} {{Hydroelastic response and energy harvesting potential of flexible piezoelectric beams in viscous flow}},\ }\bibfield  {journal} {\bibinfo  {journal} {Physics of Fluids}\ }\textbf {\bibinfo {volume} {24}},\ \href {https://doi.org/10.1063/1.4719704} {10.1063/1.4719704} (\bibinfo {year} {2012})\BibitemShut {NoStop}%
\bibitem [{\citenamefont {Michelin}\ and\ \citenamefont {Doar{\'{e}}}(2013)}]{Michelin2013EnergyFlows}%
  \BibitemOpen
  \bibfield  {author} {\bibinfo {author} {\bibfnamefont {S.}~\bibnamefont {Michelin}}\ and\ \bibinfo {author} {\bibfnamefont {D.}~\bibnamefont {Doar{\'{e}}}},\ }\bibfield  {title} {\bibinfo {title} {{Energy harvesting efficiency of piezoelectric flags in axial flows}},\ }\href {https://doi.org/10.1017/jfm.2012.494} {\bibfield  {journal} {\bibinfo  {journal} {Journal of Fluid Mechanics}\ }\textbf {\bibinfo {volume} {714}},\ \bibinfo {pages} {489} (\bibinfo {year} {2013})}\BibitemShut {NoStop}%
\bibitem [{\citenamefont {Zhang}\ \emph {et~al.}(2000)\citenamefont {Zhang}, \citenamefont {Childress}, \citenamefont {Libchaber},\ and\ \citenamefont {Shelley}}]{Zhang2000FlexibleWind}%
  \BibitemOpen
  \bibfield  {author} {\bibinfo {author} {\bibfnamefont {J.}~\bibnamefont {Zhang}}, \bibinfo {author} {\bibfnamefont {S.}~\bibnamefont {Childress}}, \bibinfo {author} {\bibfnamefont {A.}~\bibnamefont {Libchaber}},\ and\ \bibinfo {author} {\bibfnamefont {M.}~\bibnamefont {Shelley}},\ }\bibfield  {title} {\bibinfo {title} {{Flexible filaments in a flowing soap film as a model for one-dimensional flags in a two-dimensional wind}},\ }\href@noop {} {\bibfield  {journal} {\bibinfo  {journal} {Nature}\ }\textbf {\bibinfo {volume} {408}},\ \bibinfo {pages} {835} (\bibinfo {year} {2000})}\BibitemShut {NoStop}%
\bibitem [{\citenamefont {Shelley}\ \emph {et~al.}(2005)\citenamefont {Shelley}, \citenamefont {Vandenberghe},\ and\ \citenamefont {Zhang}}]{Shelley2005HeavyWater}%
  \BibitemOpen
  \bibfield  {author} {\bibinfo {author} {\bibfnamefont {M.}~\bibnamefont {Shelley}}, \bibinfo {author} {\bibfnamefont {N.}~\bibnamefont {Vandenberghe}},\ and\ \bibinfo {author} {\bibfnamefont {J.}~\bibnamefont {Zhang}},\ }\bibfield  {title} {\bibinfo {title} {{Heavy flags undergo spontaneous oscillations in flowing water}},\ }\bibfield  {journal} {\bibinfo  {journal} {Physical review letters}\ }\textbf {\bibinfo {volume} {94}},\ \href {https://doi.org/10.1103/PhysRevLett.94.094302{\"{i}}} {10.1103/PhysRevLett.94.094302{\"{i}}} (\bibinfo {year} {2005})\BibitemShut {NoStop}%
\bibitem [{\citenamefont {Shelley}\ and\ \citenamefont {Zhang}(2011)}]{Shelley2011FlappingFlows}%
  \BibitemOpen
  \bibfield  {author} {\bibinfo {author} {\bibfnamefont {M.~J.}\ \bibnamefont {Shelley}}\ and\ \bibinfo {author} {\bibfnamefont {J.}~\bibnamefont {Zhang}},\ }\bibfield  {title} {\bibinfo {title} {{Flapping and bending bodies interacting with fluid flows}},\ }\href {https://doi.org/10.1146/annurev-fluid-121108-145456} {\bibfield  {journal} {\bibinfo  {journal} {Annual Review of Fluid Mechanics}\ }\textbf {\bibinfo {volume} {43}},\ \bibinfo {pages} {449} (\bibinfo {year} {2011})}\BibitemShut {NoStop}%
\bibitem [{\citenamefont {Connell}\ and\ \citenamefont {Yue}(2007)}]{Connell2007FlappingStream}%
  \BibitemOpen
  \bibfield  {author} {\bibinfo {author} {\bibfnamefont {B.~S.}\ \bibnamefont {Connell}}\ and\ \bibinfo {author} {\bibfnamefont {D.~K.}\ \bibnamefont {Yue}},\ }\bibfield  {title} {\bibinfo {title} {{Flapping dynamics of a flag in a uniform stream}},\ }\href {https://doi.org/10.1017/S0022112007005307} {\bibfield  {journal} {\bibinfo  {journal} {Journal of Fluid Mechanics}\ }\textbf {\bibinfo {volume} {581}},\ \bibinfo {pages} {33} (\bibinfo {year} {2007})}\BibitemShut {NoStop}%
\bibitem [{\citenamefont {Jaiman}\ \emph {et~al.}(2014)\citenamefont {Jaiman}, \citenamefont {Parmar},\ and\ \citenamefont {Gurugubelli}}]{Jaiman2014AddedChannel}%
  \BibitemOpen
  \bibfield  {author} {\bibinfo {author} {\bibfnamefont {R.~K.}\ \bibnamefont {Jaiman}}, \bibinfo {author} {\bibfnamefont {M.~K.}\ \bibnamefont {Parmar}},\ and\ \bibinfo {author} {\bibfnamefont {P.~S.}\ \bibnamefont {Gurugubelli}},\ }\bibfield  {title} {\bibinfo {title} {{Added mass and aeroelastic stability of a flexible plate interacting with mean flow in a confined channel}},\ }\bibfield  {journal} {\bibinfo  {journal} {Journal of Applied Mechanics}\ }\textbf {\bibinfo {volume} {81}},\ \href {https://doi.org/10.1115/1.4025304} {10.1115/1.4025304} (\bibinfo {year} {2014})\BibitemShut {NoStop}%
\bibitem [{\citenamefont {Liu}\ \emph {et~al.}(2014)\citenamefont {Liu}, \citenamefont {Jaiman},\ and\ \citenamefont {Gurugubelli}}]{Liu2014AEffects}%
  \BibitemOpen
  \bibfield  {author} {\bibinfo {author} {\bibfnamefont {J.}~\bibnamefont {Liu}}, \bibinfo {author} {\bibfnamefont {R.~K.}\ \bibnamefont {Jaiman}},\ and\ \bibinfo {author} {\bibfnamefont {P.~S.}\ \bibnamefont {Gurugubelli}},\ }\bibfield  {title} {\bibinfo {title} {{A stable second-order scheme for fluid-structure interaction with strong added-mass effects}},\ }\href {https://doi.org/10.1016/j.jcp.2014.04.020} {\bibfield  {journal} {\bibinfo  {journal} {Journal of Computational Physics}\ }\textbf {\bibinfo {volume} {270}},\ \bibinfo {pages} {687} (\bibinfo {year} {2014})}\BibitemShut {NoStop}%
\bibitem [{\citenamefont {Eloy}\ \emph {et~al.}(2007)\citenamefont {Eloy}, \citenamefont {Souilliez},\ and\ \citenamefont {Schouveiler}}]{Eloy2007FlutterPlate}%
  \BibitemOpen
  \bibfield  {author} {\bibinfo {author} {\bibfnamefont {C.}~\bibnamefont {Eloy}}, \bibinfo {author} {\bibfnamefont {C.}~\bibnamefont {Souilliez}},\ and\ \bibinfo {author} {\bibfnamefont {L.}~\bibnamefont {Schouveiler}},\ }\bibfield  {title} {\bibinfo {title} {{Flutter of a rectangular plate}},\ }\href {https://doi.org/10.1016/j.jfluidstructs.2007.02.002} {\bibfield  {journal} {\bibinfo  {journal} {Journal of Fluids and Structures}\ }\textbf {\bibinfo {volume} {23}},\ \bibinfo {pages} {904} (\bibinfo {year} {2007})}\BibitemShut {NoStop}%
\bibitem [{\citenamefont {Eloy}\ \emph {et~al.}(2008)\citenamefont {Eloy}, \citenamefont {Lagrange}, \citenamefont {Souilliez},\ and\ \citenamefont {Schouveiler}}]{Eloy2008AeroelasticFlow}%
  \BibitemOpen
  \bibfield  {author} {\bibinfo {author} {\bibfnamefont {C.}~\bibnamefont {Eloy}}, \bibinfo {author} {\bibfnamefont {R.}~\bibnamefont {Lagrange}}, \bibinfo {author} {\bibfnamefont {C.}~\bibnamefont {Souilliez}},\ and\ \bibinfo {author} {\bibfnamefont {L.}~\bibnamefont {Schouveiler}},\ }\bibfield  {title} {\bibinfo {title} {{Aeroelastic instability of cantilevered flexible plates in uniform flow}},\ }\href {https://doi.org/10.1017/S002211200800284X} {\bibfield  {journal} {\bibinfo  {journal} {Journal of Fluid Mechanics}\ }\textbf {\bibinfo {volume} {611}},\ \bibinfo {pages} {97} (\bibinfo {year} {2008})}\BibitemShut {NoStop}%
\bibitem [{\citenamefont {Gurugubelli}\ and\ \citenamefont {Jaiman}(2015{\natexlab{a}})}]{Gurugubelli2015Self-inducedFlow}%
  \BibitemOpen
  \bibfield  {author} {\bibinfo {author} {\bibfnamefont {P.~S.}\ \bibnamefont {Gurugubelli}}\ and\ \bibinfo {author} {\bibfnamefont {R.~K.}\ \bibnamefont {Jaiman}},\ }\bibfield  {title} {\bibinfo {title} {{Self-induced flapping dynamics of a flexible inverted foil in a uniform flow}},\ }\href {https://doi.org/10.1017/jfm.2015.515} {\bibfield  {journal} {\bibinfo  {journal} {Journal of Fluid Mechanics}\ }\textbf {\bibinfo {volume} {781}},\ \bibinfo {pages} {657} (\bibinfo {year} {2015}{\natexlab{a}})}\BibitemShut {NoStop}%
\bibitem [{\citenamefont {Gurugubelli}\ and\ \citenamefont {Jaiman}(2019)}]{Gurugubelli2019LargePeriodicity}%
  \BibitemOpen
  \bibfield  {author} {\bibinfo {author} {\bibfnamefont {P.~S.}\ \bibnamefont {Gurugubelli}}\ and\ \bibinfo {author} {\bibfnamefont {R.~K.}\ \bibnamefont {Jaiman}},\ }\bibfield  {title} {\bibinfo {title} {{Large amplitude flapping of an inverted elastic foil in uniform flow with spanwise periodicity}},\ }\href {https://doi.org/10.1016/j.jfluidstructs.2019.05.009} {\bibfield  {journal} {\bibinfo  {journal} {Journal of Fluids and Structures}\ }\textbf {\bibinfo {volume} {90}},\ \bibinfo {pages} {139} (\bibinfo {year} {2019})}\BibitemShut {NoStop}%
\bibitem [{\citenamefont {Goza}\ \emph {et~al.}(2018)\citenamefont {Goza}, \citenamefont {Colonius},\ and\ \citenamefont {Sader}}]{Goza2018GlobalFlapping}%
  \BibitemOpen
  \bibfield  {author} {\bibinfo {author} {\bibfnamefont {A.}~\bibnamefont {Goza}}, \bibinfo {author} {\bibfnamefont {T.}~\bibnamefont {Colonius}},\ and\ \bibinfo {author} {\bibfnamefont {J.~E.}\ \bibnamefont {Sader}},\ }\bibfield  {title} {\bibinfo {title} {{Global modes and nonlinear analysis of inverted-flag flapping}},\ }\href {https://doi.org/10.1017/jfm.2018.728} {\bibfield  {journal} {\bibinfo  {journal} {Journal of Fluid Mechanics}\ }\textbf {\bibinfo {volume} {857}},\ \bibinfo {pages} {312} (\bibinfo {year} {2018})}\BibitemShut {NoStop}%
\bibitem [{\citenamefont {Tavallaeinejad}\ \emph {et~al.}(2018)\citenamefont {Tavallaeinejad}, \citenamefont {Legrand},\ and\ \citenamefont {Pa{\"\i}doussis}}]{Tavallaeinejad2018NonlinearFlow}%
  \BibitemOpen
  \bibfield  {author} {\bibinfo {author} {\bibfnamefont {M.}~\bibnamefont {Tavallaeinejad}}, \bibinfo {author} {\bibfnamefont {M.}~\bibnamefont {Legrand}},\ and\ \bibinfo {author} {\bibfnamefont {M.}~\bibnamefont {Pa{\"\i}doussis}},\ }\bibfield  {title} {\bibinfo {title} {Nonlinear dynamics of slender inverted flags in axial flow},\ }in\ \href@noop {} {\emph {\bibinfo {booktitle} {9th International Symposium on Fluid-Structure Interactions, Flow-Sound Interactions, Flow-Induced Vibration \& Noise}}}\ (\bibinfo {year} {2018})\BibitemShut {NoStop}%
\bibitem [{\citenamefont {Tavallaeinejad}\ \emph {et~al.}(2020)\citenamefont {Tavallaeinejad}, \citenamefont {Pa{\"\i}doussis}, \citenamefont {Legrand},\ and\ \citenamefont {Kheiri}}]{Tavallaeinejad2020InstabilityFlow}%
  \BibitemOpen
  \bibfield  {author} {\bibinfo {author} {\bibfnamefont {M.}~\bibnamefont {Tavallaeinejad}}, \bibinfo {author} {\bibfnamefont {M.~P.}\ \bibnamefont {Pa{\"\i}doussis}}, \bibinfo {author} {\bibfnamefont {M.}~\bibnamefont {Legrand}},\ and\ \bibinfo {author} {\bibfnamefont {M.}~\bibnamefont {Kheiri}},\ }\bibfield  {title} {\bibinfo {title} {Instability and the post-critical behaviour of two-dimensional inverted flags in axial flow},\ }\href@noop {} {\bibfield  {journal} {\bibinfo  {journal} {Journal of Fluid Mechanics}\ }\textbf {\bibinfo {volume} {890}},\ \bibinfo {pages} {A14} (\bibinfo {year} {2020})}\BibitemShut {NoStop}%
\bibitem [{\citenamefont {Shoele}\ and\ \citenamefont {Mittal}(2016)}]{Shoele2016EnergyFlag}%
  \BibitemOpen
  \bibfield  {author} {\bibinfo {author} {\bibfnamefont {K.}~\bibnamefont {Shoele}}\ and\ \bibinfo {author} {\bibfnamefont {R.}~\bibnamefont {Mittal}},\ }\bibfield  {title} {\bibinfo {title} {{Energy harvesting by flow-induced flutter in a simple model of an inverted piezoelectric flag}},\ }\href {https://doi.org/10.1017/jfm.2016.40} {\bibfield  {journal} {\bibinfo  {journal} {Journal of Fluid Mechanics}\ }\textbf {\bibinfo {volume} {790}},\ \bibinfo {pages} {582} (\bibinfo {year} {2016})}\BibitemShut {NoStop}%
\bibitem [{\citenamefont {Tang}\ \emph {et~al.}(2015)\citenamefont {Tang}, \citenamefont {Liu},\ and\ \citenamefont {Lu}}]{Tang2015DynamicsFlow}%
  \BibitemOpen
  \bibfield  {author} {\bibinfo {author} {\bibfnamefont {C.}~\bibnamefont {Tang}}, \bibinfo {author} {\bibfnamefont {N.~S.}\ \bibnamefont {Liu}},\ and\ \bibinfo {author} {\bibfnamefont {X.~Y.}\ \bibnamefont {Lu}},\ }\bibfield  {title} {\bibinfo {title} {{Dynamics of an inverted flexible plate in a uniform flow}},\ }\bibfield  {journal} {\bibinfo  {journal} {Physics of Fluids}\ }\textbf {\bibinfo {volume} {27}},\ \href {https://doi.org/10.1063/1.4923281} {10.1063/1.4923281} (\bibinfo {year} {2015})\BibitemShut {NoStop}%
\bibitem [{\citenamefont {Sader}\ \emph {et~al.}(2016{\natexlab{a}})\citenamefont {Sader}, \citenamefont {Coss{\'{e}}}, \citenamefont {Kim}, \citenamefont {Fan},\ and\ \citenamefont {Gharib}}]{Sader2016Large-amplitudeVibration}%
  \BibitemOpen
  \bibfield  {author} {\bibinfo {author} {\bibfnamefont {J.~E.}\ \bibnamefont {Sader}}, \bibinfo {author} {\bibfnamefont {J.}~\bibnamefont {Coss{\'{e}}}}, \bibinfo {author} {\bibfnamefont {D.}~\bibnamefont {Kim}}, \bibinfo {author} {\bibfnamefont {B.}~\bibnamefont {Fan}},\ and\ \bibinfo {author} {\bibfnamefont {M.}~\bibnamefont {Gharib}},\ }\bibfield  {title} {\bibinfo {title} {{Large-amplitude flapping of an inverted flag in a uniform steady flow-a vortex-induced vibration}},\ }\href {https://doi.org/10.1017/jfm.2016.139} {\bibfield  {journal} {\bibinfo  {journal} {Journal of Fluid Mechanics}\ }\textbf {\bibinfo {volume} {793}},\ \bibinfo {pages} {524} (\bibinfo {year} {2016}{\natexlab{a}})}\BibitemShut {NoStop}%
\bibitem [{\citenamefont {Ryu}\ \emph {et~al.}(2015)\citenamefont {Ryu}, \citenamefont {Park}, \citenamefont {Kim},\ and\ \citenamefont {Sung}}]{Ryu2015FlappingFlow}%
  \BibitemOpen
  \bibfield  {author} {\bibinfo {author} {\bibfnamefont {J.}~\bibnamefont {Ryu}}, \bibinfo {author} {\bibfnamefont {S.~G.}\ \bibnamefont {Park}}, \bibinfo {author} {\bibfnamefont {B.}~\bibnamefont {Kim}},\ and\ \bibinfo {author} {\bibfnamefont {H.~J.}\ \bibnamefont {Sung}},\ }\bibfield  {title} {\bibinfo {title} {{Flapping dynamics of an inverted flag in a uniform flow}},\ }\href {https://doi.org/10.1016/j.jfluidstructs.2015.06.006} {\bibfield  {journal} {\bibinfo  {journal} {Journal of Fluids and Structures}\ }\textbf {\bibinfo {volume} {57}},\ \bibinfo {pages} {159} (\bibinfo {year} {2015})}\BibitemShut {NoStop}%
\bibitem [{\citenamefont {Gurugubelli}\ and\ \citenamefont {Jaiman}(2015{\natexlab{b}})}]{Gurugubelli2015EnergyFoil}%
  \BibitemOpen
  \bibfield  {author} {\bibinfo {author} {\bibfnamefont {P.~S.}\ \bibnamefont {Gurugubelli}}\ and\ \bibinfo {author} {\bibfnamefont {R.~K.}\ \bibnamefont {Jaiman}},\ }\bibfield  {title} {\bibinfo {title} {{Energy Harvesting using flapping dynamics of piezoelectric inverted flexible foil}},\ }in\ \href@noop {} {\emph {\bibinfo {booktitle} {International Conference on Offshore Mechanics and Arctic Engineering}}},\ Vol.\ \bibinfo {volume} {56574},\ \bibinfo {organization} {American Society of Mechanical Engineers}\ (\bibinfo  {publisher} {American Society of Mechanical Engineers},\ \bibinfo {year} {2015})\ p.\ \bibinfo {pages} {V009T09A010}\BibitemShut {NoStop}%
\bibitem [{\citenamefont {Alam}\ \emph {et~al.}(2021)\citenamefont {Alam}, \citenamefont {Chao}, \citenamefont {Rehman}, \citenamefont {Ji},\ and\ \citenamefont {Wang}}]{Alam2021EnergyFoil}%
  \BibitemOpen
  \bibfield  {author} {\bibinfo {author} {\bibfnamefont {M.~M.}\ \bibnamefont {Alam}}, \bibinfo {author} {\bibfnamefont {L.~M.}\ \bibnamefont {Chao}}, \bibinfo {author} {\bibfnamefont {S.}~\bibnamefont {Rehman}}, \bibinfo {author} {\bibfnamefont {C.}~\bibnamefont {Ji}},\ and\ \bibinfo {author} {\bibfnamefont {H.}~\bibnamefont {Wang}},\ }\bibfield  {title} {\bibinfo {title} {{Energy harvesting from passive oscillation of inverted foil}},\ }\bibfield  {journal} {\bibinfo  {journal} {Physics of Fluids}\ }\textbf {\bibinfo {volume} {33}},\ \href {https://doi.org/10.1063/5.0056567} {10.1063/5.0056567} (\bibinfo {year} {2021})\BibitemShut {NoStop}%
\bibitem [{\citenamefont {Dabiri}(2011)}]{dabiri2011potential}%
  \BibitemOpen
  \bibfield  {author} {\bibinfo {author} {\bibfnamefont {J.~O.}\ \bibnamefont {Dabiri}},\ }\bibfield  {title} {\bibinfo {title} {Potential order-of-magnitude enhancement of wind farm power density via counter-rotating vertical-axis wind turbine arrays},\ }\href@noop {} {\bibfield  {journal} {\bibinfo  {journal} {Journal of renewable and sustainable energy}\ }\textbf {\bibinfo {volume} {3}} (\bibinfo {year} {2011})}\BibitemShut {NoStop}%
\bibitem [{\citenamefont {Hu}\ \emph {et~al.}(2020{\natexlab{a}})\citenamefont {Hu}, \citenamefont {Feng},\ and\ \citenamefont {Wang}}]{Hu2020-tandem}%
  \BibitemOpen
  \bibfield  {author} {\bibinfo {author} {\bibfnamefont {Y.~W.}\ \bibnamefont {Hu}}, \bibinfo {author} {\bibfnamefont {L.~H.}\ \bibnamefont {Feng}},\ and\ \bibinfo {author} {\bibfnamefont {J.~J.}\ \bibnamefont {Wang}},\ }\bibfield  {title} {\bibinfo {title} {Flow-structure interactions of two tandem inverted flags in a water tunnel},\ }\bibfield  {journal} {\bibinfo  {journal} {Physics of Fluids}\ }\textbf {\bibinfo {volume} {32}},\ \href {https://doi.org/10.1063/5.0012544} {10.1063/5.0012544} (\bibinfo {year} {2020}{\natexlab{a}})\BibitemShut {NoStop}%
\bibitem [{\citenamefont {Huang}\ \emph {et~al.}(2018)\citenamefont {Huang}, \citenamefont {Wei},\ and\ \citenamefont {Lu}}]{Huang2018CouplingFlow}%
  \BibitemOpen
  \bibfield  {author} {\bibinfo {author} {\bibfnamefont {H.}~\bibnamefont {Huang}}, \bibinfo {author} {\bibfnamefont {H.}~\bibnamefont {Wei}},\ and\ \bibinfo {author} {\bibfnamefont {X.~Y.}\ \bibnamefont {Lu}},\ }\bibfield  {title} {\bibinfo {title} {{Coupling performance of tandem flexible inverted flags in a uniform flow}},\ }\href {https://doi.org/10.1017/jfm.2017.875} {\bibfield  {journal} {\bibinfo  {journal} {Journal of Fluid Mechanics}\ }\textbf {\bibinfo {volume} {837}},\ \bibinfo {pages} {461} (\bibinfo {year} {2018})}\BibitemShut {NoStop}%
\bibitem [{\citenamefont {Huertas-Cerdeira}\ \emph {et~al.}(2018)\citenamefont {Huertas-Cerdeira}, \citenamefont {Fan},\ and\ \citenamefont {Gharib}}]{Huertas-Cerdeira2018CoupledFlags}%
  \BibitemOpen
  \bibfield  {author} {\bibinfo {author} {\bibfnamefont {C.}~\bibnamefont {Huertas-Cerdeira}}, \bibinfo {author} {\bibfnamefont {B.}~\bibnamefont {Fan}},\ and\ \bibinfo {author} {\bibfnamefont {M.}~\bibnamefont {Gharib}},\ }\bibfield  {title} {\bibinfo {title} {{Coupled motion of two side-by-side inverted flags}},\ }\href {https://doi.org/10.1016/j.jfluidstructs.2017.11.005} {\bibfield  {journal} {\bibinfo  {journal} {Journal of Fluids and Structures}\ }\textbf {\bibinfo {volume} {76}},\ \bibinfo {pages} {527} (\bibinfo {year} {2018})}\BibitemShut {NoStop}%
\bibitem [{\citenamefont {Hu}\ \emph {et~al.}(2020{\natexlab{b}})\citenamefont {Hu}, \citenamefont {Feng},\ and\ \citenamefont {Wang}}]{Hu2020}%
  \BibitemOpen
  \bibfield  {author} {\bibinfo {author} {\bibfnamefont {Y.~W.}\ \bibnamefont {Hu}}, \bibinfo {author} {\bibfnamefont {L.~H.}\ \bibnamefont {Feng}},\ and\ \bibinfo {author} {\bibfnamefont {J.~J.}\ \bibnamefont {Wang}},\ }\bibfield  {title} {\bibinfo {title} {Flow-structure interactions of two parallel inverted flags with small separation distances in a water tunnel},\ }\bibfield  {journal} {\bibinfo  {journal} {Journal of Fluids and Structures}\ }\textbf {\bibinfo {volume} {94}},\ \href {https://doi.org/10.1016/j.jfluidstructs.2020.102960} {10.1016/j.jfluidstructs.2020.102960} (\bibinfo {year} {2020}{\natexlab{b}})\BibitemShut {NoStop}%
\bibitem [{\citenamefont {Mazharmanesh}\ \emph {et~al.}(2020)\citenamefont {Mazharmanesh}, \citenamefont {Young}, \citenamefont {Tian},\ and\ \citenamefont {Lai}}]{Mazharmanesh2020EnergyArrangements}%
  \BibitemOpen
  \bibfield  {author} {\bibinfo {author} {\bibfnamefont {S.}~\bibnamefont {Mazharmanesh}}, \bibinfo {author} {\bibfnamefont {J.}~\bibnamefont {Young}}, \bibinfo {author} {\bibfnamefont {F.~B.}\ \bibnamefont {Tian}},\ and\ \bibinfo {author} {\bibfnamefont {J.~C.}\ \bibnamefont {Lai}},\ }\bibfield  {title} {\bibinfo {title} {{Energy harvesting of two inverted piezoelectric flags in tandem, side-by-side and staggered arrangements}},\ }\bibfield  {journal} {\bibinfo  {journal} {International Journal of Heat and Fluid Flow}\ }\textbf {\bibinfo {volume} {83}},\ \href {https://doi.org/10.1016/j.ijheatfluidflow.2020.108589} {10.1016/j.ijheatfluidflow.2020.108589} (\bibinfo {year} {2020})\BibitemShut {NoStop}%
\bibitem [{\citenamefont {Williamson}\ and\ \citenamefont {Roshko}(1988)}]{williamson1988vortex}%
  \BibitemOpen
  \bibfield  {author} {\bibinfo {author} {\bibfnamefont {C.~H.}\ \bibnamefont {Williamson}}\ and\ \bibinfo {author} {\bibfnamefont {A.}~\bibnamefont {Roshko}},\ }\bibfield  {title} {\bibinfo {title} {Vortex formation in the wake of an oscillating cylinder},\ }\href@noop {} {\bibfield  {journal} {\bibinfo  {journal} {Journal of fluids and structures}\ }\textbf {\bibinfo {volume} {2}},\ \bibinfo {pages} {355} (\bibinfo {year} {1988})}\BibitemShut {NoStop}%
\bibitem [{\citenamefont {Jaiman}\ \emph {et~al.}(2023)\citenamefont {Jaiman}, \citenamefont {Li},\ and\ \citenamefont {Chizfam}}]{Jaiman_FIV}%
  \BibitemOpen
  \bibfield  {author} {\bibinfo {author} {\bibfnamefont {R.}~\bibnamefont {Jaiman}}, \bibinfo {author} {\bibfnamefont {G.}~\bibnamefont {Li}},\ and\ \bibinfo {author} {\bibfnamefont {A.}~\bibnamefont {Chizfam}},\ }\href@noop {} {\emph {\bibinfo {title} {{Mechanics of flow-induced vibration: Physical modeling and control strategies}}}}\ (\bibinfo  {publisher} {Springer Nature},\ \bibinfo {year} {2023})\BibitemShut {NoStop}%
\bibitem [{\citenamefont {Zdravkovich}(1987)}]{zdravkovich1987effects}%
  \BibitemOpen
  \bibfield  {author} {\bibinfo {author} {\bibfnamefont {M.}~\bibnamefont {Zdravkovich}},\ }\bibfield  {title} {\bibinfo {title} {The effects of interference between circular cylinders in cross flow},\ }\href@noop {} {\bibfield  {journal} {\bibinfo  {journal} {Journal of fluids and structures}\ }\textbf {\bibinfo {volume} {1}},\ \bibinfo {pages} {239} (\bibinfo {year} {1987})}\BibitemShut {NoStop}%
\bibitem [{\citenamefont {Assi}\ \emph {et~al.}(2006)\citenamefont {Assi}, \citenamefont {Meneghini}, \citenamefont {Aranha}, \citenamefont {Bearman},\ and\ \citenamefont {Casaprima}}]{assi2006experimental}%
  \BibitemOpen
  \bibfield  {author} {\bibinfo {author} {\bibfnamefont {G.}~\bibnamefont {Assi}}, \bibinfo {author} {\bibfnamefont {J.~R.}\ \bibnamefont {Meneghini}}, \bibinfo {author} {\bibfnamefont {J.~A.~P.}\ \bibnamefont {Aranha}}, \bibinfo {author} {\bibfnamefont {P.~W.}\ \bibnamefont {Bearman}},\ and\ \bibinfo {author} {\bibfnamefont {E.}~\bibnamefont {Casaprima}},\ }\bibfield  {title} {\bibinfo {title} {Experimental investigation of flow-induced vibration interference between two circular cylinders},\ }\href@noop {} {\bibfield  {journal} {\bibinfo  {journal} {Journal of fluids and structures}\ }\textbf {\bibinfo {volume} {22}},\ \bibinfo {pages} {819} (\bibinfo {year} {2006})}\BibitemShut {NoStop}%
\bibitem [{\citenamefont {Assi}\ \emph {et~al.}(2010)\citenamefont {Assi}, \citenamefont {Bearman},\ and\ \citenamefont {Meneghini}}]{assi2010wake}%
  \BibitemOpen
  \bibfield  {author} {\bibinfo {author} {\bibfnamefont {G.~R.}\ \bibnamefont {Assi}}, \bibinfo {author} {\bibfnamefont {P.}~\bibnamefont {Bearman}},\ and\ \bibinfo {author} {\bibfnamefont {J.}~\bibnamefont {Meneghini}},\ }\bibfield  {title} {\bibinfo {title} {On the wake-induced vibration of tandem circular cylinders: the vortex interaction excitation mechanism},\ }\href@noop {} {\bibfield  {journal} {\bibinfo  {journal} {Journal of Fluid Mechanics}\ }\textbf {\bibinfo {volume} {661}},\ \bibinfo {pages} {365} (\bibinfo {year} {2010})}\BibitemShut {NoStop}%
\bibitem [{\citenamefont {Mysa}\ \emph {et~al.}(2016)\citenamefont {Mysa}, \citenamefont {Kaboudian},\ and\ \citenamefont {Jaiman}}]{mysa2016origin}%
  \BibitemOpen
  \bibfield  {author} {\bibinfo {author} {\bibfnamefont {R.~C.}\ \bibnamefont {Mysa}}, \bibinfo {author} {\bibfnamefont {A.}~\bibnamefont {Kaboudian}},\ and\ \bibinfo {author} {\bibfnamefont {R.~K.}\ \bibnamefont {Jaiman}},\ }\bibfield  {title} {\bibinfo {title} {On the origin of wake-induced vibration in two tandem circular cylinders at low {R}eynolds number},\ }\href@noop {} {\bibfield  {journal} {\bibinfo  {journal} {Journal of Fluids and Structures}\ }\textbf {\bibinfo {volume} {61}},\ \bibinfo {pages} {76} (\bibinfo {year} {2016})}\BibitemShut {NoStop}%
\bibitem [{\citenamefont {Akayd{\i}n}\ \emph {et~al.}(2010)\citenamefont {Akayd{\i}n}, \citenamefont {Elvin},\ and\ \citenamefont {Andreopoulos}}]{akaydin2010wake}%
  \BibitemOpen
  \bibfield  {author} {\bibinfo {author} {\bibfnamefont {H.}~\bibnamefont {Akayd{\i}n}}, \bibinfo {author} {\bibfnamefont {N.}~\bibnamefont {Elvin}},\ and\ \bibinfo {author} {\bibfnamefont {Y.}~\bibnamefont {Andreopoulos}},\ }\bibfield  {title} {\bibinfo {title} {Wake of a cylinder: a paradigm for energy harvesting with piezoelectric materials},\ }\href@noop {} {\bibfield  {journal} {\bibinfo  {journal} {Experiments in Fluids}\ }\textbf {\bibinfo {volume} {49}},\ \bibinfo {pages} {291} (\bibinfo {year} {2010})}\BibitemShut {NoStop}%
\bibitem [{\citenamefont {Kim}\ \emph {et~al.}(2017)\citenamefont {Kim}, \citenamefont {Kang},\ and\ \citenamefont {Kim}}]{kim2017dynamics}%
  \BibitemOpen
  \bibfield  {author} {\bibinfo {author} {\bibfnamefont {H.}~\bibnamefont {Kim}}, \bibinfo {author} {\bibfnamefont {S.}~\bibnamefont {Kang}},\ and\ \bibinfo {author} {\bibfnamefont {D.}~\bibnamefont {Kim}},\ }\bibfield  {title} {\bibinfo {title} {Dynamics of a flag behind a bluff body},\ }\href@noop {} {\bibfield  {journal} {\bibinfo  {journal} {Journal of Fluids and Structures}\ }\textbf {\bibinfo {volume} {71}},\ \bibinfo {pages} {1} (\bibinfo {year} {2017})}\BibitemShut {NoStop}%
\bibitem [{\citenamefont {Alben}(2021)}]{Alben2021CollectiveEfficiency}%
  \BibitemOpen
  \bibfield  {author} {\bibinfo {author} {\bibfnamefont {S.}~\bibnamefont {Alben}},\ }\bibfield  {title} {\bibinfo {title} {{Collective locomotion of two-dimensional lattices of flapping plates. Part 2. Lattice flows and propulsive efficiency}},\ }\bibfield  {journal} {\bibinfo  {journal} {Journal of Fluid Mechanics}\ }\textbf {\bibinfo {volume} {915}},\ \href {https://doi.org/10.1017/jfm.2021.43} {10.1017/jfm.2021.43} (\bibinfo {year} {2021})\BibitemShut {NoStop}%
\bibitem [{\citenamefont {Ojo}\ \emph {et~al.}(2022)\citenamefont {Ojo}, \citenamefont {Kohtanen}, \citenamefont {Jiang}, \citenamefont {Brody}, \citenamefont {Erturk},\ and\ \citenamefont {Shoele}}]{ojo2022flapping}%
  \BibitemOpen
  \bibfield  {author} {\bibinfo {author} {\bibfnamefont {O.}~\bibnamefont {Ojo}}, \bibinfo {author} {\bibfnamefont {E.}~\bibnamefont {Kohtanen}}, \bibinfo {author} {\bibfnamefont {A.}~\bibnamefont {Jiang}}, \bibinfo {author} {\bibfnamefont {J.}~\bibnamefont {Brody}}, \bibinfo {author} {\bibfnamefont {A.}~\bibnamefont {Erturk}},\ and\ \bibinfo {author} {\bibfnamefont {K.}~\bibnamefont {Shoele}},\ }\bibfield  {title} {\bibinfo {title} {Flapping dynamics of an inverted flag behind a cylinder},\ }\href@noop {} {\bibfield  {journal} {\bibinfo  {journal} {Bioinspiration \& Biomimetics}\ }\textbf {\bibinfo {volume} {17}},\ \bibinfo {pages} {065011} (\bibinfo {year} {2022})}\BibitemShut {NoStop}%
\bibitem [{\citenamefont {Sharman}\ \emph {et~al.}(2005)\citenamefont {Sharman}, \citenamefont {Lien}, \citenamefont {Davidson},\ and\ \citenamefont {Norberg}}]{sharman2005numerical}%
  \BibitemOpen
  \bibfield  {author} {\bibinfo {author} {\bibfnamefont {B.}~\bibnamefont {Sharman}}, \bibinfo {author} {\bibfnamefont {F.-S.}\ \bibnamefont {Lien}}, \bibinfo {author} {\bibfnamefont {L.}~\bibnamefont {Davidson}},\ and\ \bibinfo {author} {\bibfnamefont {C.}~\bibnamefont {Norberg}},\ }\bibfield  {title} {\bibinfo {title} {Numerical predictions of low {R}eynolds number flows over two tandem circular cylinders},\ }\href@noop {} {\bibfield  {journal} {\bibinfo  {journal} {International Journal for Numerical Methods in Fluids}\ }\textbf {\bibinfo {volume} {47}},\ \bibinfo {pages} {423} (\bibinfo {year} {2005})}\BibitemShut {NoStop}%
\bibitem [{\citenamefont {Sader}\ \emph {et~al.}(2016{\natexlab{b}})\citenamefont {Sader}, \citenamefont {Huertas-Cerdeira},\ and\ \citenamefont {Gharib}}]{Sader2016StabilityFlow}%
  \BibitemOpen
  \bibfield  {author} {\bibinfo {author} {\bibfnamefont {J.~E.}\ \bibnamefont {Sader}}, \bibinfo {author} {\bibfnamefont {C.}~\bibnamefont {Huertas-Cerdeira}},\ and\ \bibinfo {author} {\bibfnamefont {M.}~\bibnamefont {Gharib}},\ }\bibfield  {title} {\bibinfo {title} {{Stability of slender inverted flags and rods in uniform steady flow}},\ }\href {https://doi.org/10.1017/jfm.2016.691} {\bibfield  {journal} {\bibinfo  {journal} {Journal of Fluid Mechanics}\ }\textbf {\bibinfo {volume} {809}},\ \bibinfo {pages} {873} (\bibinfo {year} {2016}{\natexlab{b}})}\BibitemShut {NoStop}%
\bibitem [{\citenamefont {Lighthill}(1986)}]{lighthill1986informal}%
  \BibitemOpen
  \bibfield  {author} {\bibinfo {author} {\bibfnamefont {J.}~\bibnamefont {Lighthill}},\ }\href@noop {} {\emph {\bibinfo {title} {An informal introduction to theoretical fluid mechanics}}}\ (\bibinfo  {publisher} {Oxford University Press, New York, NY},\ \bibinfo {year} {1986})\BibitemShut {NoStop}%
\bibitem [{\citenamefont {Shiels}\ \emph {et~al.}(2001)\citenamefont {Shiels}, \citenamefont {Leonard},\ and\ \citenamefont {Roshko}}]{shiels2001flow}%
  \BibitemOpen
  \bibfield  {author} {\bibinfo {author} {\bibfnamefont {D.}~\bibnamefont {Shiels}}, \bibinfo {author} {\bibfnamefont {A.}~\bibnamefont {Leonard}},\ and\ \bibinfo {author} {\bibfnamefont {A.}~\bibnamefont {Roshko}},\ }\bibfield  {title} {\bibinfo {title} {Flow-induced vibration of a circular cylinder at limiting structural parameters},\ }\href@noop {} {\bibfield  {journal} {\bibinfo  {journal} {Journal of Fluids and Structures}\ }\textbf {\bibinfo {volume} {15}},\ \bibinfo {pages} {3} (\bibinfo {year} {2001})}\BibitemShut {NoStop}%
\bibitem [{\citenamefont {Gharib}\ \emph {et~al.}(1997)\citenamefont {Gharib}, \citenamefont {Shiels}, \citenamefont {Gharib}, \citenamefont {Leonard},\ and\ \citenamefont {Roshko}}]{gharib1997exploration}%
  \BibitemOpen
  \bibfield  {author} {\bibinfo {author} {\bibfnamefont {M.~R.}\ \bibnamefont {Gharib}}, \bibinfo {author} {\bibfnamefont {D.~G.}\ \bibnamefont {Shiels}}, \bibinfo {author} {\bibfnamefont {M.}~\bibnamefont {Gharib}}, \bibinfo {author} {\bibfnamefont {A.}~\bibnamefont {Leonard}},\ and\ \bibinfo {author} {\bibfnamefont {A.}~\bibnamefont {Roshko}},\ }\bibfield  {title} {\bibinfo {title} {Exploration of flow-induced vibration at low mass and damping},\ }in\ \href@noop {} {\emph {\bibinfo {booktitle} {ASME International Mechanical Engineering Congress and Exposition}}},\ Vol.\ \bibinfo {volume} {26751}\ (\bibinfo {organization} {American Society of Mechanical Engineers},\ \bibinfo {year} {1997})\ pp.\ \bibinfo {pages} {75--81}\BibitemShut {NoStop}%
\end{thebibliography}%

\end{document}